\documentclass[12pt, draftclsnofoot, onecolumn]{IEEEtran} \usepackage{lipsum}
\usepackage{amsmath}
\usepackage{graphicx}
\usepackage[export]{adjustbox}
\usepackage{tikz}
\usepackage{pgfplots}
\usepackage{amssymb}
\usepackage[caption=false]{subfig}
\usepackage[utf8]{inputenc}
\usetikzlibrary{decorations.pathreplacing}
\graphicspath{{images/}}
\usetikzlibrary{patterns}
\usepgfplotslibrary{fillbetween}
\pgfplotsset{compat=1.15,width=10cm}
\usepackage{cite}
\usepackage{optidef}
\usepackage[ruled,linesnumbered]{algorithm2e}
\usepackage{setspace}
\usepackage{multicol}

\usepackage{booktabs}
\usepackage{framed}
\usepackage{multirow}
\usepackage{float}
\makeatletter
\newcommand{\mathleft}{\@fleqntrue\@mathmargin0pt}
\newcommand{\mathcenter}{\@fleqnfalse}
\makeatother

\usepackage{comment}
\usepackage{float}
\providecommand{\NOOP}[1]{}

\title{Revenue Maximization through Cell Switching and Spectrum Leasing in 5G HetNets}

\author{Attai Ibrahim Abubakar\IEEEauthorrefmark{1}, Cihat Ozturk\IEEEauthorrefmark{2}, Metin Ozturk\IEEEauthorrefmark{2}, Michael S. Mollel\IEEEauthorrefmark{3}, 
Syed Muhammad Asad\IEEEauthorrefmark{1}, 
Naveed Ul Hassan\IEEEauthorrefmark{4}, Sajjad Hussain\IEEEauthorrefmark{1}, and Muhammad Ali Imran\IEEEauthorrefmark{1}
\thanks{\IEEEauthorrefmark{1}Communication, sensing and imaging (CSI) research group, James Watt School of Engineering, University of Glasgow, United Kingdom. 
Email: a.abubakar.1@research.gla.ac.uk, 
\IEEEauthorrefmark{2}Faculty of Engineering and Natural Sciences, Ankara Yıldırım Beyazıt University, Ankara, Turkey. 
\IEEEauthorrefmark{3}The Nelson Mandela African Institution of Science and Technology (NM-AIST), Arusha, Tanzania. 
\IEEEauthorrefmark{4}Lahore University of Management Sciences, Lahore, Pakistan. 
}
}

\begin{document}
\maketitle
\begin{abstract}
One of the ways of achieving improved capacity in mobile cellular networks is via network densification.  Even though densification increases the capacity of the network, it also leads to increased energy consumption which can be curbed by dynamically switching off some base stations~(BSs) during periods of low traffic. However, dynamic cell switching has the challenge of spectrum under-utilization as the spectrum originally occupied by the BSs that are turned off remains dormant. This dormant spectrum can be leased by the primary network~(PN) operators, who hold the license, to the secondary network~(SN) operators who cannot afford to purchase the spectrum license. Thus enabling the PN to gain additional revenue from spectrum leasing as well as from electricity cost savings due to reduced energy consumption.
Therefore, in this work, we propose a cell switching and spectrum leasing framework based on simulated annealing (SA) algorithm to maximize the revenue of the PN while respecting the quality-of-service constraints. 
The performance evaluation reveals that the proposed method is very close to optimal exhaustive search method with a significant reduction in the computation complexity.
\end{abstract}
\section{Introduction}
The demand for capacity improvement in order to achieve enhanced data transmission is a challenge that is constantly facing mobile network operators~(MNOs). 
This is due to increase in the number of connected devices, increasing use of data hungry applications, such as online gaming and multimedia services, as well as other emerging use cases including virtual and augmented reality, driver-less cars, etc~\cite{MORGADO2018}.
In addition, with the proliferation of Internet of things~(IoT) devices where virtually everything is connected to the Internet, the demand for more capacity would further escalate~\cite{Akpakwu2018}.

One of the major approaches for enhancing network capacity to meet the ever increasing data demands in 5G is the introduction of network densification~\cite{Kamel2016}, which
involves the deployment of massive number of small base stations~(SBSs), including micro remote radio head (RRH), pico, and femto, under the coverage of macro base stations~(MBSs), by employing the principle of spatial frequency reuse.
However, this comes at a cost of increased network energy consumption (as 5G also targets to be 100 times more energy efficient that 4G networks)~\cite{Buzzi2016}. Also, with 6G network targeting higher capacity and data rates due to emerging use cases, the network densification would further increase in 6G~\cite{Giordani2020}.
This means that the energy consumption of the network would further escalate if not properly managed.
To tackle the problem of increased energy consumption owing to capacity expansion through network densification, the most common approach is to implement dynamic network operation; i.e., the base stations~(BSs) are only available when needed. This can be achieved via dynamic cell switching and traffic offloading~\cite{Feng2017}. 

In dynamic cell switching approach, the BSs~(which accounts for about 50\% - 60\% of the total power consumption of the radio access networks~(RAN))~\cite{Buzzi2016} are turned off when they are not serving any user demand or have very few users connected to them, while the traffic of the BSs that are turned off are transferred to the neighbouring BSs or MBSs. This ensures that the energy consumption of the network scales with the capacity utilization thereby enhancing the energy efficiency of the network.
Dynamic cell switching has the advantage of minimizing energy consumption of the network which translates to cost savings or additional revenue on the side of the mobile network operators~(MNOs) due to reduction in the expenditure on energy purchase. 
It also results in reduced green house gas emission, as most of the energy used to power the BSs are from fossil fuels, thus ensuring environmental sustainability~\cite{ALAMU2020}.

One of the major drawbacks of dynamic cell switching is that the spectrum allocated to the switched off BSs remains dormant during the period that they are inactive, resulting in spectrum under-utilization. 
Such dormant spectrum can be leased to smaller network operators~(also known as secondary networks~(SN) operators) who require a smaller amount of spectrum for their data transmission and cannot afford to purchase a spectrum license like the major network operators~(also known as primary network~(PN) operators). This is because spectrum is normally auctioned by the telecommunication regulatory body in each country~(e.g., Office of Communications~(Ofcom) in the UK) at a very expensive rate. Spectrum leasing results in enhanced spectrum utilization and additional revenue to the PN operators, since it has been observed that the licensed spectrum is not always fully utilized most of the time~\cite{Yua2018}.
The spectrum purchased by the SN from the PN can be used to provide data services which are delay tolerant~(DT) such as meter readings, health information from wearables, etc., and do not require real-time data transmission. It can also be used to provide non-delay tolerant~(NDT) services such as location and traffic update services, voice calls, etc., which require real-time data transmission for quick decision making.  Therefore, the PN operator can gain revenue both from energy cost savings due to dynamic cell switching and from leasing the dormant spectrum of the BSs that are turned off to the SN. 

Several approaches have been proposed in literature for implementing dynamic cell switching in mobile cellular networks~\cite{Asad2019, Abubakar2020, Ozturk2021, Zhang2020, Wu2020, Amine2020, Alsafasfeh2020}. These methods comprise analytical, heuristic and machine learning based approaches. Similar optimization techniques have also been proposed for spectrum leasing~\cite{Tan2018, Tsirakis2018, Liu2018, Bilibashi2020, Liu2020, Xiao2020, Ozturk2019}.
However, very few research works have considered both cell switching and spectrum leasing for maximizing the revenue of the PN~\cite{Sboui2015} and~\cite{Sboui2016}, even though only a homogeneous network deployment scenario as well as a fixed electricity and spectrum pricing policy were considered, thereby making their work quite simplistic.

Therefore, in this paper, we propose a cell switching and spectrum leasing framework for maximizing the revenue of the PN. 
The solution to this problem is non-trivial as it involves trying different options out of a large set of possibilities. The problem becomes very difficult to solve when the number of SBSs deployed in the network becomes very large. The optimal solution is the exhaustive search~(ES) approach because it tries all the possible options before selecting the best one. However, due to the huge computational overhead involved in implementing ES, we develop a heuristic solution that can find a near optimal solution with much lesser complexity. A cell switching and spectrum leasing scheme based on simulated annealing~(SA) algorithm is developed to determine the optimal cell switching and spectrum leasing strategy that would result in maximum revenue for the PN by considering both the revenue obtained from energy savings as a result of switching off some SBSs and that obtained from leasing the spectrum to the SN. The SA algorithm also ensures that the quality of service~(QoS) of the PN is maintained by ensuring that the options which violate the QoS of the network are avoided when selecting the optimal solution. 
To the best of our knowledge, this is the first work to consider heterogeneous network~(HetNet) scenario for the PN as the previous works~\cite{Sboui2015} and~\cite{Sboui2016} considered only homogeneous networks. In addition, both fixed and dynamic electricity and spectrum pricing policies are considered in this work, thus making our work novel and more realistic.

The remaining part of this paper is organized as follows: In Section II we review the related literature, while in Section III the system model is comprehensively presented. The proposed SA algorithm based framework for cell switching and spectrum leasing is discussed in Section IV, followed by the performance evaluation in Section V. Section VI concludes the work.

\section{Related Works}

Dynamic cell switching techniques are the most commonly employed methods for optimizing energy consumption in cellular networks because they are the cheapest to implement and require minimal changes to network architecture~\cite{Buzzi2016}. These techniques result in significant energy savings compared to other methods such as cell zooming, bandwidth adaptation, sectorization, etc~\cite{Feng2017, ALAMU2020}. 
The authors in~\cite{Asad2019} proposed a reinforcement learning based cell switching approach to optimize the energy efficiency as well as the $\text{CO}_2$ emission in a HetNet.
A cell switching and traffic offloading scheme for energy optimization in ultra-dense network using artificial neural network was proposed in~\cite{Abubakar2020}. 
The authors in~\cite{Ozturk2021} developed a scalable reinforcement learning based cell switching framework using state-action-reward-state-action~(SARSA) algorithm with value function approximation to determine the optimal switching policy that would minimize the energy consumption in an ultra dense network while ensuring that the QoS of the network is maintained. 
The work in~\cite{Zhang2020} proposed deep reinforcement learning model for minimizing the energy consumption in a RAN by dynamically switching off/on some of the BSs without compromising the QoS of the network.

The authors in~\cite{Alsafasfeh2020} considered the problem of SBS power control and user association in HetNets and proposed an optimization scheme that enables SBSs to be switched off during periods of low traffic. A heuristic algorithm was developed to determine the switching pattern of redundant SBSs.
The authors in~\cite{Amine2020} proposed a delay constrained multi-level SBS switching mechanism for energy optimization in HetNets. 
A distributed $Q$-learning algorithm was developed to determine the optimal sleep level of the SBSs while considering their traffic load, interference level and volume of data in the buffer of the SBSs.
Even though dynamic cell switching results in significant energy savings, it also results in spectrum under-utilization as the spectrum that was originally allocated to the SBSs that are switched off remain dormant when they are inactive. These dormant spectrum can be exploited via spectrum leasing operations.

As regards spectrum leasing, three major reasons for spectrum leasing have been advanced in literature~\cite{vassaki2015}: i) For monetary gains, ii) to maximize transmission rates, and iii) to reduce the energy consumption of primary users~(PUs). In the first case, the PN leases some of its spectrum to the SN at a cost in order to generate additional revenue. In the second case, the PN shares some of its spectrum to the SN in exchange for assistance in data transmission, thereby enhancing the data rates of the PUs. In the third case, the secondary users~(SUs) act as a relay to the PUs thereby reducing the transmission distance between the PUs and the BSs which leads to energy savings in the PUs. 
In this work, we are interested in the first case that is spectrum leasing for monetary gains because we want to maximize the revenue of the PN. 

In this regard, various research works using techniques such as game theory, matching theory, and machine learning techniques, etc., have been proposed~\cite{Tan2018, Tsirakis2018, Liu2018, Bilibashi2020, Liu2020, Xiao2020, Ozturk2019}.  
The authors in~\cite{Tan2018} proposed a traffic-adaptive spectrum leasing scheme whereby the SUs are able to negotiate the duration of channel leasing with the PUs in order to ensure their continual utilization of the leased channel for the complete transmission of the data in their buffer. To achieve this objective, the average utilities of both the PN and SN were first formulated, after which a spectrum leasing agreement that is beneficial to both parties was developed using Stackelberg game model.
The work in~\cite{Liu2018} proposed a joint optimization scheme for spectrum leasing and spectrum allocation using both Stackelberg game and matching theory. The proposed approach is able to determine the best price for leasing the spectrum as well as the best PU-SU pair while enhancing the spectral efficiency of the PUs and SUs.
In~\cite{Tsirakis2018}, the authors considered a spectrum leasing problem between MNOs and mobile virtual network operators~(MVNOs) using matching theory in order to maximize the utilities of the both parties in terms of spectrum leasing cost and bandwidth allocation. Their goal is to find a suitable paring between the MNOs and MVNOs that would maximize the revenue of the MNOs as well as the bandwidth allocated to the MVNOs. 

The work in~\cite{Ozturk2019} considered the problem of spectrum leasing optimization for CRN transmission over TV white spaces. A neural network based solution was proposed to determine the optimal transmission policy that would result in minimal spectrum leasing cost while considering the QoS of the CRN.
The works in~\cite{Bilibashi2020} and~\cite{Liu2020} considered the problem of resource allocation and spectrum leasing in CRNs where the PUs lease part of their spectrum to the SUs in exchange for data transmission assistance from the SUs as well energy saving for the PUs. A resource optimization model for the CRN to minimize the power consumption of the PUs, while considering the uncertainty of the communication environment was proposed.
In~\cite{Xiao2020}, an adaptive spectrum leasing with channel aggregation for CRN was considered where the amount of spectrum that the PUs can lease to SUs varies with the number of active transmissions as well as the amount of buffered data. 
A leasing algorithm was developed to adjust the amount of spectrum to be leased while satisfying the requirement of both PUs and SUs.

Joint cell switching and spectrum leasing has been considered in~\cite{Sboui2015} and~\cite{Sboui2016} to maximize the profit of both PN and SN as well as to minimize the energy consumption of PN.
The authors in~\cite{Sboui2015} considered a CRN comprising both PN and SN where the PN aims to reduce its energy consumption by turning off some BSs and transferring the users to the SN to maintain their QoS. In addition, the PN obtains revenue by leasing the free spectrum to the SN while the SN also gains revenue from the PN by charging a roaming price. A sub-optimal heuristic algorithm was developed to optimize the energy consumption of the PN by determining the set of BSs to switch off per time. 

In this paper, we develop a cell switching and spectrum leasing framework for revenue maximization in a HetNet. Different from previous works~\cite{Sboui2015} and~\cite{Sboui2016} which considered only homogeneous network deployment scenario, a HetNet deployment scenario with different types of SBSs is considered in this work, which makes our work more realistic. In addition, we did not consider offloading the traffic load of the PN SBSs that are switched off to the SN as was the case in previous works, as this would lead to additional expenses on the part of the PN due to roaming charges. Rather, we consider vertical traffic offloading from the switched off SBSs to the MBS in order to maintain the QoS and maximize the profit of the PN. Furthermore, a fixed pricing policy was considered in the previous works while in this work, both fixed and dynamic electricity and spectrum pricing policies as well as DT and NDT spectrum demand scenarios are considered. This is because both pricing policies and spectrum leasing scenarios are a better representation of what is obtainable in real systems.
\subsection{Contributions}
In this paper, we propose a cell switching and spectrum leasing framework to maximize the revenue of the PN. The proposed algorithm is able to learn the optimal cell switching and spectrum leasing policy that would result in maximum revenue for the PN while ensuring that the QoS of the PN is maintained. The proposed framework is implemented  locally at each MBS since they are responsible for controlling the SBSs under their coverage.
The following are the contributions of this work:
\begin{itemize}
    \item We formulate the problem as a binary integer programming problem and develop a cell switching and spectrum leasing framework using SA algorithm to determine the optimal policy that maximizes the revenue of the PN while ensuring that the QoS is maintained.
    \item We consider a HetNet comprising four different types of SBSs, which makes the network scenario more complex and realistic compared to the previous works that considered only homogeneous scenario.
    \item We consider two electricity and spectrum pricing policies: 1) fixed, and 2) dynamic policy, in order to study the effects of constant and varying electricity and spectrum prices on the maximum revenue of the PN, as both could be the cases faced in real systems. For the dynamic pricing policy, both DT and NDT spectrum demand scenarios are also investigated.
    \item In addition to the ES algorithm, two benchmark solutions are also developed for comparison with the proposed framework.
    \item A complexity comparison of the proposed method with that of the ES is carried out to highlight the advantage of the proposed framework.
    \item Finally, in order to capture the realistic behaviour of the network, the performance of the proposed framework is evaluated using real data comprising call detail records~(CDR) of Milan city via extensive simulations and the result obtained is compared with benchmarks.
\end{itemize}

\section{System Model}
\subsection{Network Model}
Two types of networks are considered: First, the PN is a HetNet with control and data separated architecture~(CDSA)~\cite{Mohamed2016} consisting of MBSs and SBSs. The MBSs serve as control BSs, provide constant coverage and low data rate transmission. The SBSs are deployed within the coverage of the MBSs and serve as data BSs to provide high data rate transmissions in hot spot zones. The communication between MBS and SBSs are carried out in the control channels, which are separated from the data channels.
Four types of SBSs---RRH, micro, pico and femto---are considered.
Second, a SN is also assumed to operate in the same coverage area and the HetNet allows this SN to lease some unused spectrum whenever PN SBSs are put to sleep.

\begin{figure}[h!]
\centering
\includegraphics[width=0.4\columnwidth]{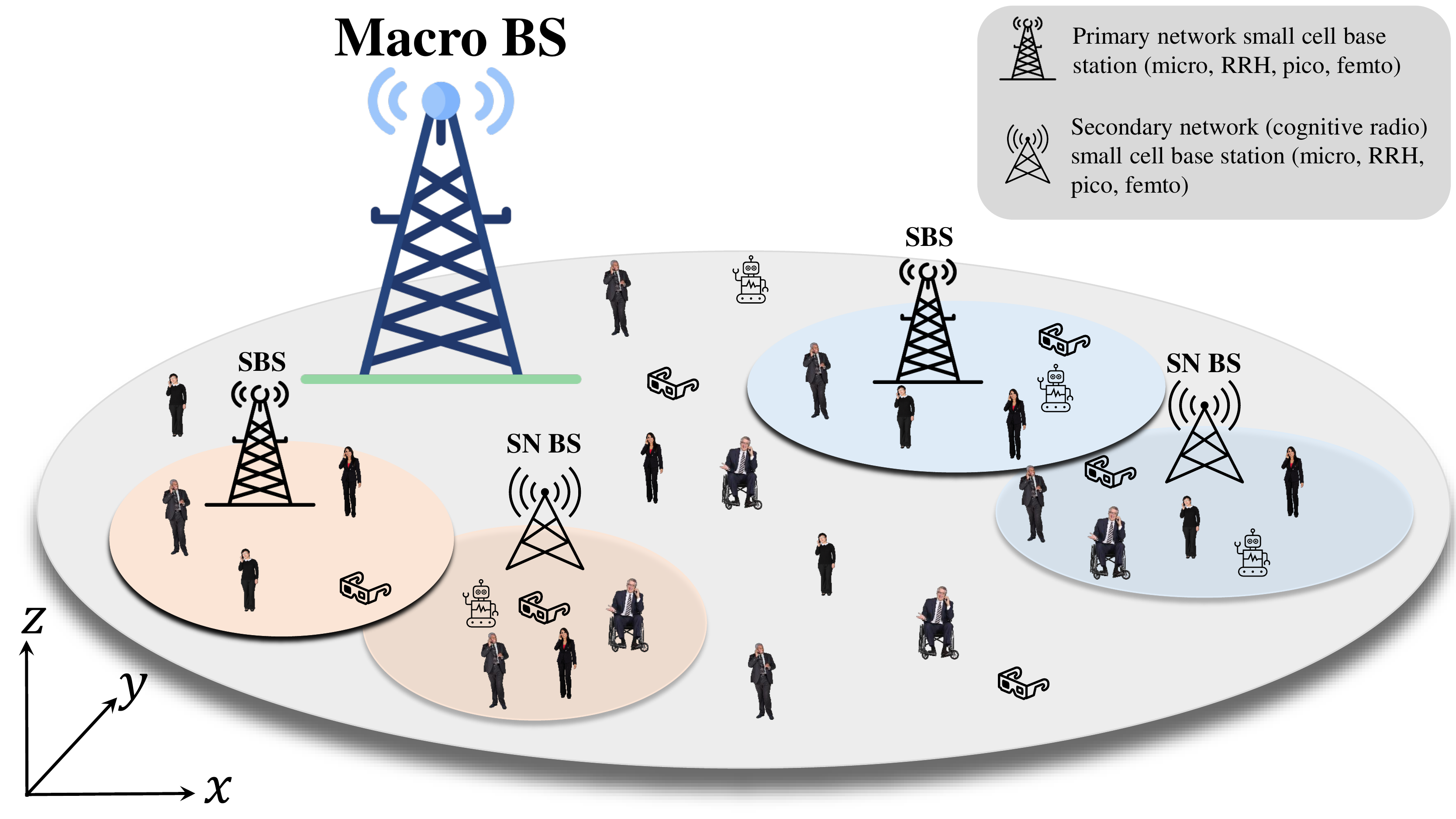}
\caption{The PN comprises a HetNet deployment of MBS and various types of SBSs and the SN comprises SN BSs.}\label{fig:sys_model}
\end{figure}

In this context, it is assumed that each SBS continuously monitors the activity of the SN BSs in their coverage area and reports the spectrum demand alongside their own traffic load to the MBS. The MBS then decides which set of SBSs to switch off in order to maximize the revenue of the PN based on the available capacity in the MBS, the traffic loads of the SBSs, and the spectrum demanded by the SN's BSs without violating the Quality of Service~(QoS) of the PN. 
The QoS in this work refers to the ability of the network to serve all the traffic demands placed on it. This can be maintained by ensuring that the traffic load of the SBSs that are switched off are transferred to the MBS, and that the capacity of the MBS are not exceeded during traffic offloading.
The network model is presented in Fig.~\ref{fig:sys_model}.

\subsection{Power Consumption of HetNet}
We adopt the BS power consumption model in~\cite{Auer2011, Debaillie2015} for estimating the power consumption of the BSs in the HetNet. The total power consumption of the HetNet comprises sum of the power consumption of the MBS and that of all the SBSs under its coverage.
The instantaneous power consumption of a BS, $P_\text{BS}$, at time $t$ can be expressed as:
\begin{equation}\label{BS_pwr}
  P_\text{BS}~(\tau, t) = P_\text{o} + \tau_t \zeta P_\text{tx},
\end{equation}
where $P_\text{o}$ is the constant circuit power consumption, $\tau_t$ is the instantaneous traffic load at time $t$, $\zeta$ is the load dependent power consumption component and $P_\text{tx}$ is the transmission power of the BS. It should be noted that the value of $P_\text{o}$, $\zeta$, and $P_\text{tx}$ is different for each type of BS~(i.e., MBS, RRH, micro, pico, and femto).

As such, the total power consumption of the HetNet, $P_\text{HN}$, can be expressed as:
\begin{equation}
    P_\text{HN}~(\tau, t) =
    \sum_{i} \sum_{j} P_{\text{BS}_{i,j}}(\tau, t),
\end{equation}
where $P_{\text{BS}_{i,j}}$ denotes the power consumption of the $j^{th}$ BS in the $i^{th}$ macro cell~(MC), and $P_{\text{BS}_{i,1}}$ represents power consumption of the MBS in the $i^{th}$ MC.
\subsection{Pricing Policy}
Two kinds of pricing policies are considered for both the electricity and spectrum:
\subsubsection{Fixed Pricing Policy}
The unit cost of electricity as well as that of the spectrum remains constant throughout the day, irrespective of the fluctuations in energy or spectrum demand.
\subsubsection{Dynamic Pricing Policy}
\begin{figure}[h!]
	\centering
    {\resizebox{0.34\columnwidth}{!}{
%
%
\definecolor{mycolor1}{rgb}{0.00000,0.44700,0.74100}%
\definecolor{mycolor2}{rgb}{0.85000,0.32500,0.09800}%
\begin{tikzpicture}

\begin{axis}[%
width=4.521in,
height=3.566in,
at={(0.758in,0.481in)},
label style={font=\Large},
scale only axis,
xmin=0,
xmax=144,
xlabel={Time (mins)},
ymin=0.7,
ymax=1,
ylabel={Price (Pounds)},
axis background/.style={fill=white},
xmajorgrids,
xminorgrids,
ymajorgrids,
yminorgrids,
legend style={at={(0.03,0.97)}, anchor=north west, legend cell align=left, align=left, draw=white!15!black}
]
\addplot [color=mycolor1, line width = 2.0pt]
  table[row sep=crcr]{%
1	0.831204629458062\\
2	0.826435426568896\\
3	0.823145540743931\\
4	0.81843834098\\
5	0.8156656387234\\
6	0.808826073978734\\
7	0.80481454927994\\
8	0.798006141105704\\
9	0.792079882395748\\
10	0.78694974381381\\
11	0.782178835838701\\
12	0.776314425246167\\
13	0.777618350969858\\
14	0.774814724654547\\
15	0.770468848102923\\
16	0.770599752201075\\
17	0.768909624511077\\
18	0.762482303045317\\
19	0.759900725422254\\
20	0.759246669954931\\
21	0.756001038860602\\
22	0.753750557926293\\
23	0.755202826127109\\
24	0.759875846668253\\
25	0.761373454654397\\
26	0.76508294662939\\
27	0.768444136048376\\
28	0.771243344641014\\
29	0.772227334238492\\
30	0.776968480713489\\
31	0.782683851293739\\
32	0.790351622783731\\
33	0.797937550148413\\
34	0.809090439803955\\
35	0.821760700941808\\
36	0.836910932081246\\
37	0.850994864474456\\
38	0.86539199015401\\
39	0.87791925658429\\
40	0.889925929274317\\
41	0.901795652561481\\
42	0.911521617793646\\
43	0.923844273910652\\
44	0.933160786004547\\
45	0.936393861466017\\
46	0.939486267337109\\
47	0.945824536814251\\
48	0.94587545688085\\
49	0.947938765880856\\
50	0.953136100349768\\
51	0.951976796915688\\
52	0.950342007014962\\
53	0.951630819476877\\
54	0.951580131921998\\
55	0.953680642797144\\
56	0.959931255355985\\
57	0.961842269179633\\
58	0.966172334934321\\
59	0.968754610092543\\
60	0.969327286457995\\
61	0.966340440907615\\
62	0.965660344127694\\
63	0.963172468727642\\
64	0.959866617097928\\
65	0.960810847191349\\
66	0.961088931208028\\
67	0.964194822758859\\
68	0.964587302541597\\
69	0.970283374649118\\
70	0.970939057698477\\
71	0.976790215122992\\
72	0.976437029820872\\
73	0.983040362658393\\
74	0.984107823963219\\
75	0.982580686988683\\
76	0.979471540273777\\
77	0.980550859676304\\
78	0.973629450806328\\
79	0.969541197240055\\
80	0.96929845500476\\
81	0.969388669551977\\
82	0.967172832863894\\
83	0.966994496374937\\
84	0.966375550177279\\
85	0.965655926405021\\
86	0.96278952192541\\
87	0.963321276228206\\
88	0.964815396538555\\
89	0.968056144886772\\
90	0.971571722087593\\
91	0.974081221077571\\
92	0.974857810221138\\
93	0.972427132704116\\
94	0.971226209672221\\
95	0.968427698614742\\
96	0.970165026183825\\
97	0.971589625490005\\
98	0.974087498894001\\
99	0.975412583184178\\
100	0.974385113895128\\
101	0.96898991195281\\
102	0.965206713762694\\
103	0.964052293074726\\
104	0.958465501475431\\
105	0.957794705164296\\
106	0.957459190752868\\
107	0.960081225415147\\
108	0.958311811228755\\
109	0.959017019274396\\
110	0.955904152371976\\
111	0.954189378439697\\
112	0.949504499800851\\
113	0.948060369510223\\
114	0.947218677085159\\
115	0.95133924978046\\
116	0.949208512381761\\
117	0.947169617112317\\
118	0.943213430202795\\
119	0.943236681374758\\
120	0.938291157098206\\
121	0.937685231556847\\
122	0.935587743334056\\
123	0.937253224781773\\
124	0.933677659557287\\
125	0.934665834365719\\
126	0.934605148806895\\
127	0.936257609598313\\
128	0.934201275949896\\
129	0.933221471563371\\
130	0.9270273593524\\
131	0.920091069732368\\
132	0.914322221456602\\
133	0.909245095546739\\
134	0.904031717769172\\
135	0.897768317065752\\
136	0.893461269971306\\
137	0.888904970313417\\
138	0.882253507549932\\
139	0.874073512741593\\
140	0.870683026845933\\
141	0.86297828599252\\
142	0.853750593375523\\
143	0.8463252541052\\
144	0.837552199403988\\
};
\addlegendentry{Electricity price}

\addplot [color=mycolor2, line width = 2.0pt]
  table[row sep=crcr]{%
1	0.804348320159327\\
2	0.801165449628244\\
3	0.798199921236682\\
4	0.795994800494361\\
5	0.794160916514682\\
6	0.792327032535003\\
7	0.782613219774709\\
8	0.777915164521823\\
9	0.775802425998444\\
10	0.773783541664136\\
11	0.769523613765316\\
12	0.764213469503164\\
13	0.757433807094863\\
14	0.75234907304668\\
15	0.737503094469588\\
16	0.730473583141141\\
17	0.72697412432952\\
18	0.726702421995143\\
19	0.722816701060789\\
20	0.716300722402868\\
21	0.71268382534907\\
22	0.711584211322192\\
23	0.711773617549578\\
24	0.712272733136246\\
25	0.712814321409588\\
26	0.713475008040068\\
27	0.715531011868777\\
28	0.719214765618329\\
29	0.720594980904578\\
30	0.719675606687289\\
31	0.723608522900786\\
32	0.730253655119916\\
33	0.735853824114991\\
34	0.740391059814767\\
35	0.743004429192476\\
36	0.74456504154638\\
37	0.749211175353107\\
38	0.756745297334134\\
39	0.769666047775413\\
40	0.784739839336495\\
41	0.793028377005323\\
42	0.797927022292533\\
43	0.801500923419121\\
44	0.804519065700619\\
45	0.811182020818103\\
46	0.81875641172757\\
47	0.822449729537472\\
48	0.823945750038167\\
49	0.827794867135529\\
50	0.837020144382716\\
51	0.847446672452459\\
52	0.859773389478737\\
53	0.868477781910206\\
54	0.872762422542929\\
55	0.876183326771991\\
56	0.879922119545897\\
57	0.888064848820661\\
58	0.893667875478261\\
59	0.898664837816146\\
60	0.902769350927409\\
61	0.906207196384752\\
62	0.909179743363502\\
63	0.912960432957058\\
64	0.917648863617113\\
65	0.920153777280774\\
66	0.922291410306718\\
67	0.922291410306718\\
68	0.923072867885333\\
69	0.924879964406149\\
70	0.92718197454037\\
71	0.930350720901173\\
72	0.933326387500792\\
73	0.935787472109652\\
74	0.935787472109652\\
75	0.933986710397567\\
76	0.921064747617958\\
77	0.915761274224359\\
78	0.909277987428368\\
79	0.905271258704848\\
80	0.903800142858736\\
81	0.902140816894428\\
82	0.899694310075258\\
83	0.897394308764533\\
84	0.895192930927733\\
85	0.894303498174694\\
86	0.892014607335109\\
87	0.884156835292454\\
88	0.877818247962609\\
89	0.87132200814058\\
90	0.867201919390073\\
91	0.867201919390073\\
92	0.867279640478642\\
93	0.868131211801631\\
94	0.869718528811859\\
95	0.871975521607977\\
96	0.875105486204289\\
97	0.877868495550665\\
98	0.879478616726836\\
99	0.88002761953694\\
100	0.88615955113355\\
101	0.899483613273831\\
102	0.91155291221933\\
103	0.93310895221439\\
104	0.937860426469928\\
105	0.941606933216536\\
106	0.948431204510616\\
107	0.960414843132725\\
108	0.976854568694828\\
109	0.991771201663576\\
110	0.996324665958777\\
111	0.99823374143711\\
112	0.99823374143711\\
113	0.99823374143711\\
114	0.99823374143711\\
115	0.998966296179588\\
116	1\\
117	0.998335754341522\\
118	0.996612325579718\\
119	0.991762630154559\\
120	0.970280518178311\\
121	0.955494587682141\\
122	0.940333967734553\\
123	0.930189627257421\\
124	0.925198797890827\\
125	0.914740331950468\\
126	0.908095964887127\\
127	0.901724210305295\\
128	0.896095359109278\\
129	0.893893981272478\\
130	0.893454616842244\\
131	0.894588288122507\\
132	0.89944908751482\\
133	0.902682732640358\\
134	0.905026601243506\\
135	0.906738207597513\\
136	0.907348177675262\\
137	0.90795814775301\\
138	0.908336818122157\\
139	0.907548354991926\\
140	0.905740735862479\\
141	0.90297716944436\\
142	0.900516226206984\\
143	0.89996668752298\\
144	0.899417148838977\\
};
\addlegendentry{Spectrum price}

\end{axis}
\end{tikzpicture}%
}}
    \caption{Dynamic electricity and spectrum pricing policy~(normalized) for every 10 minutes over a 24 hours period.}
    \label{fig:dyn_pri}
\end{figure}
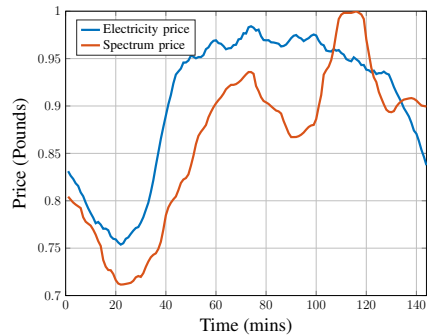

The electricity and spectrum price varies according to the amount of electricity and spectrum demanded at different times of the day. 
The dynamic pricing model for electricity was adapted from~\cite{EID2016elect_Pri}, where the instantaneous electricity prices were obtained by multiplying the fixed price by a variable factor to indicate changes in the prices at different times of the day. For the dynamic spectrum price, we assumed that the spectrum prices follow the traffic demand pattern of the PN. However, these values are scaled with the fixed spectrum price such that: $C_{\text{RB},t} = m\, . \, C_\text{RB,F}$, where $m$ is a time variable function that changes  with the instantaneous traffic load, $\tau_t$, i.e., $m = f(\tau_t)$,  $C_{\text{RB},t}$ and $C_\text{RB,F}$ are the dynamic and fixed spectrum price~(i.e., cost per resource block~(RB)). According to 3GPP~\cite{3gpp2018technical}, a RB is equivalent to 12 successive subcarriers, thus taking one subcarrier to be 15kHz, we consider one RB to be 180kHz.
The dynamic electricity and spectrum pricing policies are presented in Fig.~\ref{fig:dyn_pri}.

Under the dynamic spectrum pricing policy, two types of spectrum demand scenarios are considered:
\begin{itemize}
    \item \textbf{Non-Delay Tolerant~(NDT):} This scenario deals with applications such as location updates, voice calls, etc., that require real-time data transmission and cannot tolerate delay because of the sensitivity of the information and its requirement for quick decision making. For such applications, the SN has to demand for the spectrum as soon as the need for data transmission arises, irrespective of the spectrum price.
    \item \textbf{Delay Tolerant~(DT):} There are some other applications such as meter readings, feedback from wearables, etc., whose information may not be needed for real-time decision making and hence can tolerate some level of delay in data transmission. In these scenarios, the SN can decide to accumulate their service demands until the periods of the day where the spectrum price is cheapest, before transmission to save cost. In this work, the cheapest period is statistically decided only once and then the traffic is adjusted accordingly.
\end{itemize}

\section{Problem formulation}
A specific time period $T$ (in mins) is considered and it is then divided into equal time slots (in mins) with a duration of $d$ (in mins). Then, we define an index vector $\nu$ that stores the time slots in an order, such that $\nu=[1,2,...,L]$, where $L$ is the number of time slots and is given by $L=T/d$.
The BSs of the PN are represented by $B^P_{i,j}$ while that of the SN by $B^S_{i,j}$.
We view the problem from the PN perspective and formulate the revenue maximization problem by considering the revenue obtained from the combination of cell switching and spectrum leasing.
Since the PN obtains its power supply from the grid, it can decide to turn off some SBSs during periods of low traffic to reduce their energy cost~(i.e., gain some revenue from energy saving) and also lease the dormant spectrum to the SN in order to gain additional revenue. 
\subsubsection{Revenue from cell switching}
The power consumption when no cell switching is implemented~(i.e., when all the BSs are on), $P_\text{on}$, can be expressed as:
    \begin{equation}\label{eqn:P_ON}
    \text{P}_{\text{on}}(\tau, t)=\sum_{\nu} \sum_{i} \sum_{j} P_{\text{BS}_{i,j}}(\tau, t).
    \end{equation}
The power consumption when cell switching is implemented~(i.e., when some BSs are turned off), $\text{P}_{\text{cs}}$, is given by:
    \begin{equation}\label{eqn:P_CS}
    \text{P}_{\text{cs}}(\tau, \Gamma) = \sum_{\nu} \sum_{i} \sum_{j} [\Gamma P_{\text{BS}_{i,j}}(\tau, t) + (1 - \Gamma) P^\text{s}_{\text{BS}_{i,j}}]
    \end{equation}
where $P^\text{s}_{\text{BS}_{i,j}}$ is the power consumption of the BS when it is switched off~(i.e., sleep mode power consumption of the BSs)
$\Gamma$ denotes the on/off status of $(i,j)^{th}$ BS at time $t$ i,e.,
    \begin{equation}
    \Gamma_{i, j}=
    \begin{dcases}
    1 , &\text{if } B^P_{i,j}~\text{is on}\\
    0 , &\text{if } B^P_{i,j}~\text{is off},
    \end{dcases}
    \end{equation}
Since the MBS is always on, $\Gamma_{i, 1} = 1$, $\forall t$.

Then, the power saving due to cell switching, $P_\text{sv}$ can be expressed as:
    \begin{equation}
       P_\text{sv}(\tau, \Gamma) = P_\text{on}(\tau) - P_\text{cs}(\tau, \Gamma).
    \end{equation}
Therefore, the revenue due to energy saving,~$R_\text{E}$, can be expressed as:
\begin{equation}\label{eqn:rev_es}
        R_\text{E}(\tau, \Gamma) = \sum_\nu P_{\text{sv},t}\frac{T}{L}\text{C}_{\text{e},t},
\end{equation}
where $\text{C}_{\text{e},t}$ is the cost of electricity at any given time, $t$.
\subsubsection{Revenue from spectrum leasing}    
The revenue due to spectrum leasing,~($R_\text{l}$), can be expressed as:
    \begin{equation}\label{eqn:rev_sl_1}
    R_{\text{l}}(\tau, \Gamma)=\sum_{\nu} \sum_{i} \sum_{j} (1 - \Gamma) \min(\Psi^S_{i,j}, \Psi^D_{i,j}) \text{C}_{\text{RB},t}
    \end{equation}
where $\Psi^S_{i,j}$ denotes the amount of spectrum~(number of RBs) supplied by $B^P_{i,j}$, $\Psi^D_{i,j}$ denotes the amount of spectrum demanded by $\Psi^S_{i,j}$ from $B^P_{i,j}$ and $\text{C}_{\text{RB},t}$ is the unit cost of spectrum~(i.e., price per RB).

$B^P_{i,j}$ and $B^S_{i,j}$ are assumed to have the same capacity, which implies that $\Psi^D_{i,j} \leq \Psi^S_{i,j}$. 
Therefore,~\eqref{eqn:rev_sl_1} can be simplified as:
    \begin{equation}\label{eqn:rev_sl_2}
    R_{\text{l}}(\tau, \Gamma)=\sum_{\nu} \sum_{i} \sum_{j} (1 - \Gamma) \Psi^D_{i,j} \text{C}_{\text{RB},t}.
    \end{equation}
\subsubsection{Total Revenue} The total revenue of the PN, $R_\text{T}$ can be expressed as: 
\begin{equation}\label{eqn:Rev_T}
    R_\text{T}(\tau, \Gamma) = R_\text{E}(\tau, \Gamma) + R_{\text{l}}(\tau, \Gamma),
\end{equation}
and substituting~\eqref{eqn:rev_es} and~\eqref{eqn:rev_sl_2} in~\eqref{eqn:Rev_T}, we obtain:
\begin{equation}\label{eqn:Rev_T_2}
    R_\text{T}(\tau, \Gamma) = \sum_\nu P_{\text{sv},t}\frac{T}{L}\text{C}_{\text{e},t} + 
    \sum_{\nu} \sum_{i} \sum_{j} (1 - \Gamma) \Psi^D_{i,j} \text{C}_{\text{RB},t},
\end{equation}
Replacing~\eqref{eqn:rev_es} with~\eqref{eqn:P_ON} and~\eqref{eqn:P_CS} and simplifying~\eqref{eqn:Rev_T_2}, we get the closed form expression for the total revenue and is expressed as:
\begin{equation}\label{eqn:R_T}
R_\text{T}(\tau, \Gamma) =
\sum_{\nu} \sum_{i} \sum_{j} (1 - \Gamma) \Big[\sum_{\nu}(P_{\text{BS}_{i,j}} - P^\text{s}_{\text{BS}_{i,j}})\frac{T}{L}C_{\text{e},t} - \Psi^D_{i,j}C_{\text{RB},t}\Big].
\end{equation}

\subsubsection{Optimization objective} The revenue maximization objective function is the joint optimization of the revenue due to cell switching and spectrum leasing and can be expressed as:
\begin{align}\label{eq:max_rev}
\underset{\Gamma_{}}{\text{max.}}\quad
& R_\text{T}(\tau, \Gamma)\\
\text{s.t.}\quad
&\Upsilon_{i} = \hat T_i,\label{eq:max_QoS}\\ 
& \hat\tau_{i,1} \leq \tau^\text{m}_{i,1},\label{eq:max_MC}\\
& \Gamma \in \{0,1\}.
\end{align}
The constraints of~\eqref{eq:max_rev} are explained in the following. The traffic demand, $\Upsilon_{i}$, when all the BSs in the MC $i$ are on~(i.e., before traffic offloading) is computed as,
\begin{equation}\label{eqn:constraint1}
\Upsilon_{i} = \sum_j \tau_{i,j}.
\end{equation}
To ensure QoS, the traffic of any SBS that is switched off will be transferred to MBS and therefore the actual traffic of MBS during the offloading process, denoted by $\hat \tau_{i,1}$ is equal to,
\begin{equation}
\hat \tau_{i,1} =\tau_{i,1} + \sum_{j=2} \tau_{i,j}(1 - \Gamma_{i,j}).
\end{equation}
The traffic demand of the MC after traffic offloading, $\hat T_i$ can be expressed as,  
\begin{equation}\label{eqn:constraint2}
\hat T_i =\hat \tau_{i,1} + \sum_{j=2} \tau_{i,j}\Gamma_{i,j}.
\end{equation}
Therefore,~\eqref{eqn:constraint1} must be equal to~\eqref{eqn:constraint2} to satisfy the constraint in~\eqref{eq:max_QoS}. 
We are assuming that MBS will be able to handle the traffic of all the switched off SBSs. On the other hand, if there is a maximum limit on the amount of traffic that the MBS can handle then we also have to introduce another constraint. For example, let $\tau^\text{m}_{i,1}$ denote the maximum traffic that MBS can serve in any time slot $t$. Then, we have the additional constraint in~\eqref{eq:max_MC}. 

The solution to the problem in~\eqref{eq:max_rev} is non-trivial as it involves deciding the optimal set of SBSs to turn off out of all the possible options, and then leasing their spectrum to the SN BSs in order to maximize the revenue of the PN. The optimal solution can be obtained from ES algorithm, however, the number of search spaces increases exponentially with the number of SBSs in ES, thereby resulting in huge computational overhead. 
Hence, we resort to a less complex heuristic which considers a lesser search space and can give a  sub-optimal solution with reduced computational complexity~(lesser search spaces compared to ES). 

\section{Proposed Framework}
The aim of this paper is to determine the optimal cell switching and spectrum leasing strategy that would maximize the revenue of the PN without compromising the QoS of the network.
Although ES always finds the optimal policy, it is computationally complex to implement because it has to sequentially search through all the possible cell switching and spectrum leasing combinations before deciding the optimal solution. As a result, in this work, we employ the SA algorithm which has lesser complexity since it involves lesser search spaces in finding the optimal solution. However, this algorithm is not always guaranteed to produce the optimal result as is the case with ES. In this regard, albeit being sub-optimal, through extensive simulations, we prove that the developed SA algorithm based solution produces almost the same results as the ES algorithm---especially when the network sizes are reasonable---with much less computational complexity, providing a promising trade-off between the performance and complexity.

\subsection{Simulated Annealing~(SA) Algorithm}
The SA algorithm is a probability-based heuristic that deals with the annealing process in solid materials. 
The working principle of SA algorithm involves mimicking the process during which a heated solid material cools down. 
It is used in the optimization of difficult problems such as machine scheduling, inventory control and vehicle routing problems in the literature~\cite{wei2018simulated}.
One of the most important features of the SA algorithm is that it ensures the results that degrade the value of the objective function are included in the solution process under certain conditions in order not to be stuck at a local optimum. In other words, an improved objective function value (better than the current best solution) is always accepted, whereas non-improved solutions are accepted based on a probability value~\cite{Lin2018}.
This mechanism is elaborated in the next section.

In basic SA algorithm, the criteria used to accept a worse objective function are the random numbers between 0 and 1, the improvement in the objective function, and the current temperature values.
In the operation steps of the algorithm, as the temperature of the system decreases, the possibility of accepting worse results decreases because, as can be seen in \eqref{eqn:rev_sa}, the decrease in $\mathcal{T}$ value also decreases the selection probability of worse solutions. Thus, while the diversification feature is high at the beginning of the algorithm, intensification feature becomes prominent towards the last iterations. In other words, the algorithm performs a wider search by taking into account the worse solutions in the initial stage. However, it focuses on specific regions in the search space in the final stages.
The probability of increase at temperature, $\mathcal{T}$~(in Kelvin), of $\delta E$ amplitude in energy is presented in~\eqref{eqn:SAA}, where $K$ is Boltzman constant:
\begin{equation}\label{eqn:SAA}
    p(\delta) = \exp\Bigg(-\frac{\delta E}{K\mathcal{T}}\Bigg).
\end{equation}
Therefore, the starting temperature of the system is a hyper-parameter of the algorithm and has a significant impact on the overall performance, such that it must be high enough to allow any feasible solution to be accepted. However, if it is set too high, the search process will be random until the temperature decreases to a certain level. As such, a certain $\mathcal{T}$ value is determined as a stopping criterion in order not to prolong the search process excessively.
\subsection{SA algorithm for cell switching and spectrum leasing}
To control the switching off/on of SBSs, it is necessary to determine the parameters of the algorithm in the first place. Then the objective function value of randomly generated initial solution $s$ is calculated with~\eqref{eq:max_rev}.
In this way, the revenues are obtained according to the energy saved from turning off some SBSs in the PN~\eqref{eqn:rev_es} and spectrum leased to the SN~\eqref{eqn:rev_sl_2}. 
During the search process, the algorithm attempts to transform the current solution $s$ into one of its randomly selected new solution $s^\prime$. However, in the developed algorithm, instead of randomly selecting a neighborhood structure, each neighborhood is applied in an order as in sequential variable neighborhood search~(VNS) algorithm~\cite{hansen2010variable}. We also expanded the search area in each iteration due to the small number of neighborhood types. 

Note that only feasible solutions which guarantee~\eqref{eq:max_QoS} and~\eqref{eq:max_MC} are considered in the proposed SA algorithm. To ensure this, a feasibility check is performed first in each of the neighborhood solution produced. 
With the applied neighborhood structure, several temporal solutions can be produced until a feasible solution is obtained. 
If the revenue of the obtained solution with the new neighborhood structure, $s^\prime$, is higher than the current solution $s$, the new solution is unconditionally accepted.
If the revenue of the neighborhood solution is less than the existing solution, the probability of accepting the neighborhood solution is calculated as:
\begin{equation}\label{eqn:rev_sa}
    p = \exp\Bigg(-\frac{R_\text{T}(s^\prime) - R_\text{T}(s)}{\mathcal{T}}\Bigg).
\end{equation}
After the local search process (after $k$ iteration), the temperature is decreased according to the formula $\mathcal{T} = \mathcal{T}- \alpha$, where $\alpha$ is the temperature reduction parameter. The pseudo code for the developed SA based cell switching and spectrum leasing framework is presented in Algorithm~\ref{alg:SAA_revmax}.

The step-by-step implementation procedure of the proposed SA based cell switching and spectrum leasing framework is discussed in the following:
\subsubsection{Feasibility Check}
In order for a solution to be evaluated within the algorithm, a preliminary check is performed to determine whether it is feasible or not. For this reason, the transferred traffic loads of SBSs that are switched off in the $s$ solution should not exceed the normalized capacity of the MBS~\eqref{eq:max_MC}. The pseudo code for feasibility check is shown in Algorithm~\ref{alg:Fes_chk}.
\begin{singlespace}
\begin{algorithm}[]\label{alg:Fes_chk}
\relsize{-1.5}
\SetAlgoLined
\SetKwInOut{Input}{Input}
\SetKwInOut{Output}{Output}
\SetKwRepeat{Do}{do}{while}%
	Randomly generate an initial solution: $s_0\in S$\\
	\lWhile{$s_0$ is infeasible;}{Randomly generate an initial solution: $s_0 \in S$}
	Calculate revenue of $s_0$\\
	Define an initial temperature $\mathcal{T}>0$\\
    Define temperature reduction function and $\alpha$ value
    $s=s_0, s^*=s_0, f(s) = f(s_0), f(s^*) = f(s_0);$\\
    Define local search iteration number for each temperature~($k$)\\
    \While{$\mathcal{T}>0.01$}{
    $n = k;$\\
        \While{(n$>$0)}{
        generate~(\textbf{1-reserve}) neighbor solution $s^\prime$\\
            \While{($s^\prime$ is infeasible)}{
            generate~(\textbf{1-reserve}) neighbor solution $s^\prime$\\
            $\Delta = f(s^\prime) - f(s);$\\
            \lIf{$(\Delta \leq 0)$}{$s = s^\prime$}
             \uElse{generate a random number from uniform distribution in the 0-1 range~($u$)\\
              \lIf{$\big(u<\exp(-\frac{\Delta}{T})\big);$}{$s=s^\prime$}
             }
 	          \lIf{$(f(s^\prime) < f(s^*));$}{$s^* = s^\prime$}
            }
            
        generate~(\textbf{2-reserve}) neighbor solution $s^\prime$\\
            \While{($s^\prime$ is infeasible)}{
          generate~(\textbf{2-reserve}) neighbor solution $s^\prime$\\
            $\Delta = f(s^\prime) - f(s);$\\
            \lIf{$\tau_{i,1} \leq 1$}{$s = s^\prime$}
	         \uElse{generate a random number from uniform distribution in the 0-1 range~(u)\\
 	        \lIf{$\big(u<\exp(-\frac{\Delta}{T})\big);$}{$s=s^\prime$}
 	        }
 	        \lIf{$(f(s^\prime) < f(s^*));$}{$s^* = s^\prime$}
            }
            
        generate~(\textbf{swap}) neighbor solution $s^\prime$\\
            \While{($s^\prime$ is infeasible)}{
                	generate~(\textbf{swap}) neighbor solution $s^\prime$\\
	                $\Delta = f(s^\prime) - f(s);$\\
	                \lIf{$(\Delta \leq 0);$}{$s = s^\prime$}
	                \uElse{generate a random number from uniform distribution in the 0-1 range~($u$)
	                \lIf{$\big(u<\exp(-\frac{\Delta}{T})\big);$}{$s=s^\prime$}
	                }
	                \lIf{$(f(s^\prime) < f(s^*));$}{$s^* = s^\prime$}
            }
        $n = n - 1;$\\
        }
    	$\mathcal{T} = \mathcal{T} -\alpha$\\
	apply~(\textbf{shaking}) procedures to $s^*$, 
	$s = s^*$ \\
    }
$s^*$ is the heuristic solution of the problem
\caption{SA algorithm for cell switching and spectrum leasing}
\label{alg:SAA_revmax}
\end{algorithm}
\end{singlespace}

\begin{singlespace}
\begin{algorithm}[]\label{alg:Fes_chk}
\relsize{-1.5}
\SetAlgoLined
\SetKwInOut{Input}{Input}
\SetKwInOut{Output}{Output}
\SetKwRepeat{Do}{do}{while}%
    MBS traffic load = $\tau_\text{i,1}$\;
	\For{$i \quad \textbf{in} \quad s^\prime$}{
	    \If{($s^\prime(i) = 0$)}{
	        calculate the transferred traffic load $\sum_{j=2} \tau_{i,j}(1 -  \Gamma_{i,j})$\;
	        $\hat \tau_{i,1} =\tau_{i,1} + \sum_{j=2} \tau_{i,j}(1 - \Gamma_{i,j})$\;
	        
	            \eIf{$\hat \tau_{i,1} \leq 1$}{
	                $s^\prime$ is feasible}
	            {
	                $s^\prime$ is not feasible
	        }
	    }
	}
\caption{Feasibility check}
\label{alg:FC_revmax}
\end{algorithm}
\end{singlespace}

\subsubsection{Solution Representation}
The proposed SA algorithm has a representation scheme specially designed for the cell switching and spectrum leasing problem. It has a binary representation depending on whether the SBSs are off or on.


\subsubsection{Initial Solution}
In the SA algorithm, the initial solution, which is the first feasible solution at the beginning of the iterations in the SA algorithm, is generated randomly or with certain methodical approaches such as nearest neighbor heuristics~\cite{YuNN}. Simple heuristic methods are considered to decrease the solution time and increase the quality of the solution in some NP-hard problems. However, in this work, the initial solution is generated randomly, and not with any constructive heuristic method because the optimized initial solution can be trapped in a particular local optimum within the search space.

\subsubsection{Neighborhood Structures}
The proposed SA algorithm has three different neighborhood structures, seeking for better results from different aspects in each iteration. The SA algorithm also has nested iterations. The primary iteration is associated with temperature drop. Each temperature level represents one iteration and performs a global search in the search space. In addition, there are local search iterations in which neighborhood structures are applied sequentially at each temperature level. Neighborhood structures are named as \textbf{1-reserve}, \textbf{2-reserve} and \textbf{swap}, and they are frequently used in applications such as vehicle routing problems, travelling salesman problems~(TSP), and location problems~\cite{alvarez2018iterated, wei2018simulated}. In the neighborhood of 1-reserve, a random cell is chosen from the solution state $s$ and the selected cell's index is denoted by $j$. If the value of $s(j)$ is 1, this value is changed to 0 and vice versa for the case where the value of $s(j)$ is 0. In the 2-reserve neighborhood, this process is performed for two different cells, while in the swap neighborhood, the values of two randomly selected cells are replaced with each other. 

\begin{figure}[h!]
\centering
\resizebox{0.3\columnwidth}{!}{\begin{tikzpicture}[xscale=0.9, yscale=0.5]
\draw[line width=0.25mm] (0,0) -- ++(1,0) -- ++(0,1) -- ++(-1,0) -- cycle;
\draw[line width=0.25mm] (1,0) -- ++(1,0) -- ++(0,1) -- ++(-1,0) -- cycle;
\draw[line width=0.25mm] (2,0) -- ++(1,0) -- ++(0,1) -- ++(-1,0) -- cycle;
\draw[line width=0.25mm] (3,0) -- ++(1,0) -- ++(0,1) -- ++(-1,0) -- cycle;
\draw[line width=0.25mm] (4,0) -- ++(1,0) -- ++(0,1) -- ++(-1,0) -- cycle;
\draw[line width=0.25mm] (5,0) -- ++(1,0) -- ++(0,1) -- ++(-1,0) -- cycle;
\draw[line width=0.25mm] (6,0) -- ++(1,0) -- ++(0,1) -- ++(-1,0) -- cycle;
\draw[line width=0.25mm] (7,0) -- ++(1,0) -- ++(0,1) -- ++(-1,0) -- cycle;
\node (a) at (0.5,0.5) {1};
\node (b) at (1.5,0.5) {0};
\node (c) at (2.5,0.5) {0};
\node (d) at (3.5,0.5) {1};
\node (e) at (4.5,0.5) {1};
\node (f) at (5.5,0.5) {0};
\node (g) at (6.5,0.5) {0};
\node (h) at (7.5,0.5) {1};
\node (aa) at (7.0,-0.5) {1-reserve};

\draw[line width=0.25mm] (0,-2) -- ++(1,0) -- ++(0,1) -- ++(-1,0) -- cycle;
\draw[line width=0.25mm] (1,-2) -- ++(1,0) -- ++(0,1) -- ++(-1,0) -- cycle;
\draw[line width=0.25mm] (2,-2) -- ++(1,0) -- ++(0,1) -- ++(-1,0) -- cycle;
\draw[line width=0.25mm] (3,-2) -- ++(1,0) -- ++(0,1) -- ++(-1,0) -- cycle;
\draw[line width=0.25mm] (4,-2) -- ++(1,0) -- ++(0,1) -- ++(-1,0) -- cycle;
\draw[line width=0.25mm] (5,-2) -- ++(1,0) -- ++(0,1) -- ++(-1,0) -- cycle;
\draw[line width=0.25mm] (6,-2) -- ++(1,0) -- ++(0,1) -- ++(-1,0) -- cycle;
\draw[line width=0.25mm] (7,-2) -- ++(1,0) -- ++(0,1) -- ++(-1,0) -- cycle;
\node (a1) at (0.5,-1.5) {1};
\node (b1) at (1.5,-1.5) {1};
\node (c1) at (2.5,-1.5) {0};
\node (d1) at (3.5,-1.5) {1};
\node (e1) at (4.5,-1.5) {1};
\node (f1) at (5.5,-1.5) {0};
\node (g1) at (6.5,-1.5) {0};
\node (h1) at (7.5,-1.5) {1};
\node (aa) at (7.0,-2.5) {2-reserve};

\draw[line width=0.25mm] (0,-4) -- ++(1,0) -- ++(0,1) -- ++(-1,0) -- cycle;
\draw[line width=0.25mm] (1,-4) -- ++(1,0) -- ++(0,1) -- ++(-1,0) -- cycle;
\draw[line width=0.25mm] (2,-4) -- ++(1,0) -- ++(0,1) -- ++(-1,0) -- cycle;
\draw[line width=0.25mm] (3,-4) -- ++(1,0) -- ++(0,1) -- ++(-1,0) -- cycle;
\draw[line width=0.25mm] (4,-4) -- ++(1,0) -- ++(0,1) -- ++(-1,0) -- cycle;
\draw[line width=0.25mm] (5,-4) -- ++(1,0) -- ++(0,1) -- ++(-1,0) -- cycle;
\draw[line width=0.25mm] (6,-4) -- ++(1,0) -- ++(0,1) -- ++(-1,0) -- cycle;
\draw[line width=0.25mm] (7,-4) -- ++(1,0) -- ++(0,1) -- ++(-1,0) -- cycle;
\node (a2) at (0.5,-3.5) {1};
\node (b2) at (1.5,-3.5) {1};
\node (c2) at (2.5,-3.5) {1};
\node (d2) at (3.5,-3.5) {1};
\node (e2) at (4.5,-3.5) {0};
\node (f2) at (5.5,-3.5) {0};
\node (g2) at (6.5,-3.5) {0};
\node (h2) at (7.5,-3.5) {1};
\node (aa) at (7.0,-4.6)  {swap};

\draw[line width=0.25mm] (0,-6) -- ++(1,0) -- ++(0,1) -- ++(-1,0) -- cycle;
\draw[line width=0.25mm] (1,-6) -- ++(1,0) -- ++(0,1) -- ++(-1,0) -- cycle;
\draw[line width=0.25mm] (2,-6) -- ++(1,0) -- ++(0,1) -- ++(-1,0) -- cycle;
\draw[line width=0.25mm] (3,-6) -- ++(1,0) -- ++(0,1) -- ++(-1,0) -- cycle;
\draw[line width=0.25mm] (4,-6) -- ++(1,0) -- ++(0,1) -- ++(-1,0) -- cycle;
\draw[line width=0.25mm] (5,-6) -- ++(1,0) -- ++(0,1) -- ++(-1,0) -- cycle;
\draw[line width=0.25mm] (6,-6) -- ++(1,0) -- ++(0,1) -- ++(-1,0) -- cycle;
\draw[line width=0.25mm] (7,-6) -- ++(1,0) -- ++(0,1) -- ++(-1,0) -- cycle;
\node (a3) at (0.5,-5.5) {1};
\node (b3) at (1.5,-5.5) {0};
\node (c3) at (2.5,-5.5) {1};
\node (d3) at (3.5,-5.5) {1};
\node (e3) at (4.5,-5.5) {0};
\node (f3) at (5.5,-5.5) {1};
\node (g3) at (6.5,-5.5) {0};
\node (h3) at (7.5,-5.5) {1};
\node (aa) at (7.0,-6.5) {shaking};

\draw[line width=0.25mm] (0,-8) -- ++(1,0) -- ++(0,1) -- ++(-1,0) -- cycle;
\draw[line width=0.25mm] (1,-8) -- ++(1,0) -- ++(0,1) -- ++(-1,0) -- cycle;
\draw[line width=0.25mm] (2,-8) -- ++(1,0) -- ++(0,1) -- ++(-1,0) -- cycle;
\draw[line width=0.25mm] (3,-8) -- ++(1,0) -- ++(0,1) -- ++(-1,0) -- cycle;
\draw[line width=0.25mm] (4,-8) -- ++(1,0) -- ++(0,1) -- ++(-1,0) -- cycle;
\draw[line width=0.25mm] (5,-8) -- ++(1,0) -- ++(0,1) -- ++(-1,0) -- cycle;
\draw[line width=0.25mm] (6,-8) -- ++(1,0) -- ++(0,1) -- ++(-1,0) -- cycle;
\draw[line width=0.25mm] (7,-8) -- ++(1,0) -- ++(0,1) -- ++(-1,0) -- cycle;
\node (a4) at (0.5,-7.5) {0};
\node (b4) at (1.5,-7.5) {1};
\node (c4) at (2.5,-7.5) {0};
\node (d4) at (3.5,-7.5) {1};
\node (e4) at (4.5,-7.5) {0};
\node (f4) at (5.5,-7.5) {0};
\node (g4) at (6.5,-7.5) {0};
\node (h4) at (7.5,-7.5) {0};


\draw [red, line width=0.5mm, ->] (1.5, -0.05) -- (1.5, -0.95);
\draw [red, line width=0.5mm, ->] (2.5, -2.05) -- (2.5, -2.95 );
\draw [red, line width=0.5mm, ->] (4.5, -2.05) -- (4.5, -2.95 );
\draw[<->, red, line width=0.4mm] (1.5,-4.1) .. controls (2.5,-5) and (4.5, -5) .. (5.5,-4.1);
\end{tikzpicture}}
\caption{Illustration of the different types of neighborhood structures. The topmost bar shows the initial status of the SBSs, followed by the implementation of the three neighbourhood structures while the last bar represents the shaking operation.} 
\label{fig:ngbr_rep}
\end{figure}
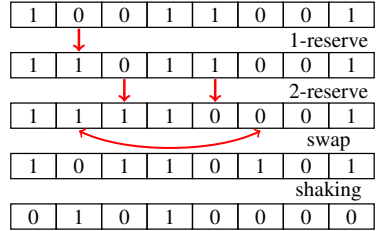

In addition to the neighborhood structures, the shaking tool is also used for diversification before each temperature change in the algorithm. After the local search procedure at certain temperature, the bit representation~(i.e., 0 and 1 values) are changed randomly to search in different spaces. This action is to prevent the algorithm from being stuck at a local optimum. The demonstration of the implementation of neighborhood structures is shown in the Fig.~\ref{fig:ngbr_rep}.

\subsubsection{Parameter Settings}
SA algorithm begins with five parameters: $\mathcal{T}$, $\mathcal{T}_F$, $k$, $\alpha$ and $K$. $\mathcal{T}$ and $\mathcal{T}_F$ are the initial and final temperatures, respectively. The initial temperature must be high enough to allow the acceptance of any feasible solution. If the initial temperature is too high, the probability of generating random solutions among feasible solutions at the beginning of the algorithm is higher. On the other hand, if the initial temperature is too low, the probability of getting stuck at the local optimum of the algorithm increases. The final temperature of the algorithm is set to avoid spending too much time in reaching the optimum. $k$ is defined as the number of iterations of the local search procedure at each temperature, while $\alpha$ is the temperature reduction parameter. It refers to the amount by which the temperature will be decayed at the end of each iteration. $K$ is Boltzmann constant and is used in calculating the probability of accepting or rejecting worse solutions. If the new objective function value is worse than current best solution, it will generate $u$, which is a random variable between 0 and 1. Then, the obtained solution will be accepted if the criterion represented in~\eqref{eqn:rev_sa} is satisfied. Except for this situation, an improved objective function value is always accepted. For more detailed information about the working mechanism of the SA algorithm, refer to~\cite{chibante2010simulated}.  
We considered different SA algorithm design parameters that are frequently used in the literature~\cite{alvarez2018iterated, wei2018simulated} and chose the ones that lead to the best results during the preliminary tests. The best SA parameter combination is $\mathcal{T} = 1$, $\alpha = 0.01$, $\mathcal{T}_F = 0.01$, $k = 10N$, where $N$ indicates the total number of SBSs in the PN. 

\subsubsection{Complexity Comparison between SA and ES}
An ES algorithm would perform a complete space search of all the possible configurations until the optimum configuration is found. This may be suitable for functions of few variables, but considering the cell switching and spectrum leasing problem, it would result in exponential computational complexity of $\mathcal{O}(2^N)$.  Due to the computational complexity of problems like this and other NP-hard problems, many optimization heuristics have been developed in order to obtain optimal or approximate optimal solutions. In addition, the solution times of heuristic approaches are incomparably low compared to algorithms that try all possible scenarios. Because, not all feasible solution combinations are considered in heuristic approaches. Heuristic approaches work with the best solution-oriented search and they focus on specific regions in the search space. Therefore, the computational cost of heuristic approaches are very low compared to ES, especially in large-scale cell switching problems. One widely used technique is the SA algorithm, by which we introduce a degree of stochasticity, potentially shifting from an optimal to a sub-optimal solution, in an attempt to reduce the complexity, escape local minima, and converge to a value closer to the global optimum. 

\begin{figure}[ht]
\centering
\resizebox{0.4\columnwidth}{!}{\definecolor{mycolor1}{rgb}{0.00000,0.44700,0.74100}%
\definecolor{mycolor2}{rgb}{0.00000,0.44706,0.74118}%
\definecolor{mycolor3}{rgb}{0.85000,0.32500,0.09800}%
\begin{tikzpicture}
\pgfplotsset{
    scale only axis,
    xmin=0, xmax=64,
    xtick={4, 8, 12, 16, 20, 24,  36,  48,  64},
      label style={font=\small},
ticklabel style={font=\small},
}
\pgfplotsset{every axis y label/.append style={mycolor1},
             every y tick label/.append style={mycolor1},
             y axis  line style={mycolor1}}
\begin{axis}[
  axis y line*=right,
  grid=both,
  ymin=0, ymax=3100,
  xlabel=Number of SBSs,
  ylabel= SA Time~(secs),
label style={font=\small},
ticklabel style={font=\large},
]
\addplot [color=mycolor1, line width=2.0pt, mark=o, mark options={solid, mycolor1}]
  coordinates{
    (4, 0.1088)
    (8, 0.4775)
    (12, 3.2501)
    (16, 10.2584)
    (20, 49.249)
    (24, 173.6565)
    (36, 426.7889)
    (48, 1239.3237)
    (64, 3049.2559)
}; \label{plot_one}


\end{axis}

\pgfplotsset{every axis y label/.append style={mycolor3},
             every y tick label/.append style={mycolor3},
             y axis  line style={mycolor3}}
\begin{axis}[
  axis y line*=left,
  axis x line=none,
  ymin=0, ymax=1100,
  ytick={  0,  175,  358,  530,  710, 890, 1100},
  ylabel= ES Time~(secs),
  legend style={at={(0.825,0.001)}, anchor=south west, legend cell align=left, align=left},
  label style={font=\small},
ticklabel style={font=\large},
legend style={font=\large}
]
\addlegendimage{color=mycolor1, line width=2.0pt, mark=o,mark options={solid, mycolor1}}
\addlegendentry{SA}
\addplot [color=mycolor3, line width=2.0pt, mark=diamond, mark options={solid, mycolor3}]
  coordinates{
    (4, 0.0414)
    (8, 0.5172)
    (12, 4.3528)
    (16, 66.6065)
    (20, 1034.4942)
}; \addlegendentry{ES}
\end{axis}

\end{tikzpicture}}
\caption{Time complexity comparison between ES and SA.}
\label{fig:time_comp}
\end{figure}
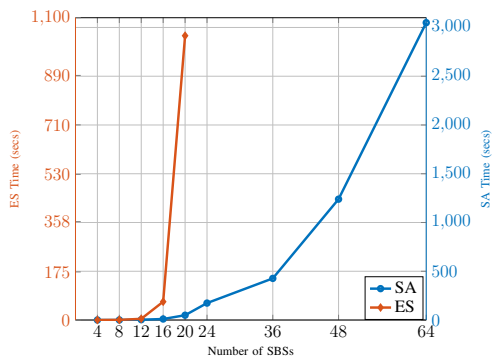

However, the time complexity of heuristic algorithms such as the SA algorithm cannot be easily determined because such algorithms do not guarantee to find the global optimal solution within a certain time limit.
Instead, determining the total simulation run time of the algorithm can give us an idea of the computational complexity of the algorithm.
Fig.~\ref{fig:time_comp} shows the simulation run time comparison between ES and the proposed SA algorithm. It can be clearly seen that the simulation run time of the ES algorithm is very small when the number of SBSs are less than 16. However, we notice a huge leap in simulation time when the number of SBSs is increased from 16 to 20 because the number of search spaces of the ES increases exponentially with the number of SBSs. This accounts for the very wide difference in the simulation time that is observed when the number of SBSs are increased to 20 compared to when they were 16. It should be noted that we stopped the simulation at 20 SBSs for the ES algorithm because of the limitation of our computer as it would take days to complete the simulation when the number of SBSs are increased to 24. The simulation time of the SA algorithm is also very low until about 20 SBS when it starts to increases with higher magnitudes. But this is much lesser than the magnitude of simulation time increase that is observed with the ES algorithm. The SA algorithm exhibits a polynomial order of computation complexity because it does not have to consider all the search spaces like the ES algorithm in order to determine the optimal cell switching and spectrum leasing strategy. Hence, the ES algorithm is only suitable for for small networks with few SBSs while the SA algorithm can be applied even when number of SBSs are very many.
\section{Performance Evaluation}
The proposed cell switching and spectrum leasing framework can be implemented in any network regardless of the network size in terms of the number of MBSs involved. Since the framework is implemented independently at each MBS, which is responsible for controlling all the SBSs under its coverage, the simulations are conducted for a single MBS with multiple SBSs for the sake of brevity.  Hence, we need to develop one framework and implement it in all the other MBS-SBSs configuration throughout the network. 
The PN, SN, and SA algorithm parameters used in the simulations are presented in Table~\ref{tab:Simulation Parameters}. 
\begin{table}[!htb]
\small
\renewcommand{\arraystretch}{.5}
\centering
\caption{Simulation Parameters}
\label{tab:Simulation Parameters}
\begin{tabular}{@{}ll@{}}
\toprule
\textbf{Parameter}                 & \textbf{Value}           \\ \midrule
Bandwidth of MBS~(MHz)                    & 20                   \\ 
Bandwidth of SBSs, SN-BSs~(MHz)           & 15, 10, 5, 3 \\ 
Number of RBs per MBS               & 100                      \\ 
Number of RBs per SBSs, SN-BSs      & 75, 50, 25, 15           \\ 
$P_\text{tx}$~(MBS, RRH, micro, pico, femto)~(W) & 20, 20, 6.3, 0.13, 0.05  \\ 
$P_\text{o}$~(MBS, RRH, micro, pico, femto) ~(W) & 130, 84, 56, 6.8, 4.8    \\ 
$\zeta$~(MBS, RRH, micro, pico, femto) & 4.7, 2.8, 2.6, 4.0, 8.0  \\ 
$P^\text{s}_{\text{BS}_{i,j}}$~(RRH, micro, pico, femto)~(W)         & 56, 39, 4.3, 2.9         \\ 
Initial temperature, $\mathcal{T}$             & 1                        \\
Final temperature, $\mathcal{T}_\text{F}$            & 0.01                     \\ 
Fixed spectrum price (per RB)      & \pounds0.13                \\ 
Fixed electricity price (per kWhr) & \pounds0.1293              \\ \bottomrule
\end{tabular}
\renewcommand{\arraystretch}{1}
\end{table}

\subsection{Data Set and Pre-processing}
To compute the total revenue of the HetNet using~\eqref{eqn:R_T}, the traffic demand of each BS in the PN~($\tau$) and SN~($\Psi$) is required. We leveraged the call detail record~(CDR) data set of the city of Milan, Italy that was made available by Telecom Italia~\cite{barlacchi2015multi}. In the data set, Milan city was divided into 10,000 square grids with each having an area of 235$\times$235 square meters. In addition, the call, short-message and Internet activities that were carried out in each grid was recorded every 10 minutes over a period of two months~(November-December 2013). Although the activity levels contained in the data set are without unit and no additional information was provided regarding how the data set was processed, we decided to interpret the CDR of each grid as the traffic loads as they signify the amount of interaction between the users and the mobile network within the grid in each time slot. However, during the data processing stage of this work, we considered only the Internet activity level as the traffic load for the PN since it was the most significant part of the data set and also considering the fact that 5G networks would be mainly Internet based. The Internet activity level of two grids were selected at random to represent the traffic load of the MBS while that of one grid was chosen for each SBS. Then, the traffic loads were normalized separately according to the capacity of each type of SBS. We assume that the traffic demand of each BS in the SN is a fraction of the traffic demand of the SBSs in the PN such that $\Psi = \beta \tau$ where $\beta$ is a variable between 0 and 1~($\beta$ was chosen to be 0.7 in this work). The traffic demand of the SN is shifted so that its maximum traffic demand coincides with the period of the day when the spectrum leasing price is minimum in order to depict the DT case while for the NDT case, the traffic demand remains intact.
\subsection{Benchmarks}
We compare the performance of the proposed method with three benchmark methods namely: ES, A-type, and D-type algorithms, which are briefly described in the following paragraphs.
\subsubsection{Exhaustive search~(ES)}
This method sequentially considers all the possible cell switching and spectrum leasing combinations in order to determine the optimal off/on switching policy that would result in maximum revenue to the PN while ensuring that the QoS of the network is maintained. Therefore, this methods is guaranteed to always find the optimal policy without violating the QoS of the network. However, the computational complexity involved in sequentially searching through all the possible combinations makes it unsuitable for online implementation. The goal of any other algorithm is to closely approximate the policy obtained from this approach, hence, it is suitable as a benchmark for this problem. 
\subsubsection{Sorting-based Algorithms}Two additional benchmark algorithms are developed using the sorting approach which we have named A-type and and D-type heuristic respectively.
In the D-type heuristic, we first evaluate a utility function, $\mathcal{N}$, which is the difference between the traffic demand of the SN BSs and that of the PN BSs, i.e., $\mathcal{N} = \Psi - \tau$. This utility function $\mathcal{N}$ is important because we are not only interested in switching off the SBSs with low traffic demand, but also those whose associated SN BS has high spectrum demand. It is necessary to satisfy both conditions if the revenue of the PN is maximized because both of them affects the total revenue~\eqref{eqn:Rev_T} that can be generated by the PN. 
In addition, since the total revenue of the PN is dependent on the amount of revenue that can be obtained from energy savings and spectrum leasing, thus, higher values of $\mathcal{N}$ would result in greater revenue generation due to higher contributions from both components. On the other hand, lower values of $\mathcal{N}$ might result in lesser revenue generation due to smaller contribution either from the energy savings or spectrum leasing. After evaluating $\mathcal{N}$, the SBSs are arranged in descending order according to the value of $\mathcal{N}$. Then, the traffic load of the SBSs are sequential offloaded to the MBS until the capacity of the MBS is reached. The procedure for implementing A-type heuristic is similar to that of D-type except that in A-type, the SBSs are sorted in ascending order according to $\mathcal{N}$. 
\subsection{Performance Metrics}
The metrics that would be used in evaluating the performance of the proposed and benchmark methods are briefly discussed in this section.
\subsubsection{Total Revenue}
The goal of this work is to determine the maximum revenue that can be obtained by the PN over a given period of time, $T$. As described in Section IV, this is obtained by combining the revenue due to energy saving from cell switching and the revenue obtained from leasing the spectrum to the SN. The total revenue of the network can be obtained from~\eqref{eq:max_rev}.
\subsubsection{Average Network Throughput}
The effect of the proposed framework on the QoS of the network is evaluated using the network throughput metric. Here, we consider the network throughput to be the traffic demand that can be served by all the remaining BSs~(both MBS and active SBSs) after cell switching and spectrum leasing operation has been executed. To estimate this throughput, we assume that the activity level contained in the employed data set are throughput demands~(in Mbps) so that $\tau$ can be seen as the normalized throughput of each BS~(i.e., MBS and SBSs). Therefore, the average network throughput $\mathcal{Q}_\text{Net}(t)$ can be obtained by aggregating the throughput demands of all active SBSs and the MBS~\cite{Ozturk2021}, such that:

\begin{equation}
    \mathcal{Q}_\text{Net}(t) = \mathcal{Q}_\text{i,1}(t) + \sum_{j=2} \mathcal{Q}_\text{i,j}(t),
\end{equation}
where $\mathcal{Q}(t)$ is the average received throughput from each BS and can be expressed as: 
\begin{equation}
    \mathcal{Q}(t)  = r_\text{m}(t)U_\text{m}(t),
\end{equation}
where $r_\text{m}$ is the average throughput that is allocated to each user~(assuming equal resource allocation) and $U_\text{m}$ is the total number of users served by each BS at a given time $t$.

\subsection{Results and Discussions}
		

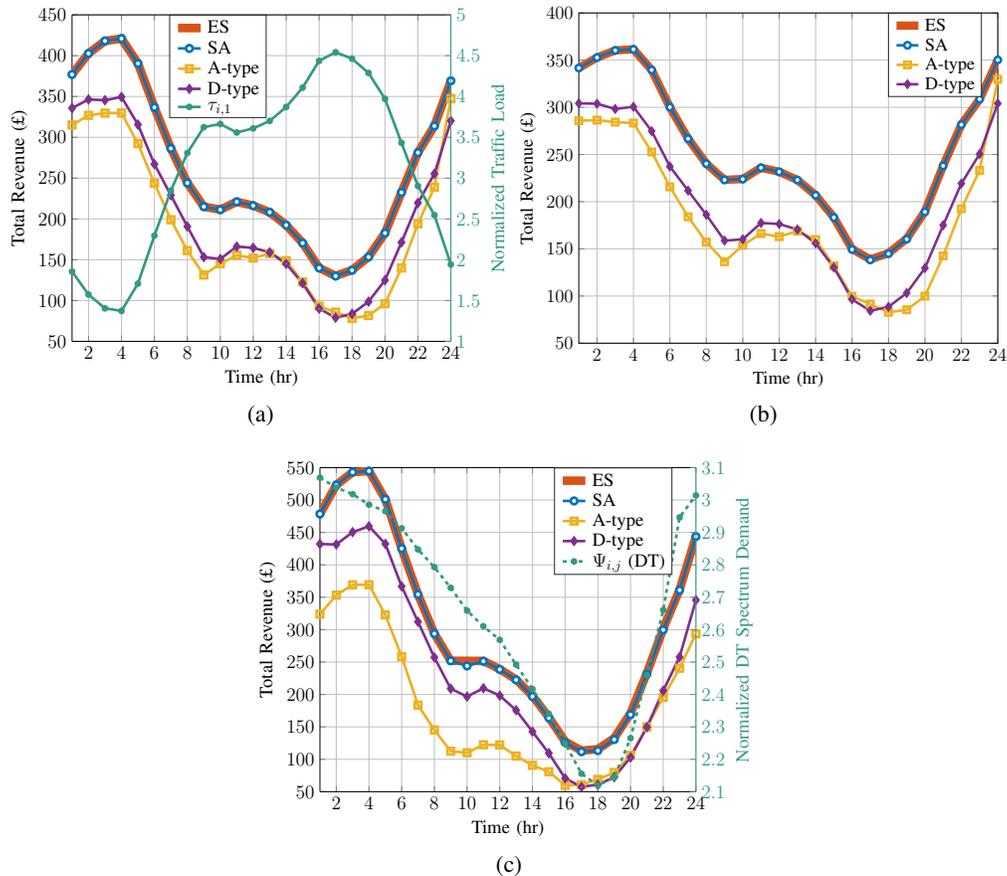
\begin{figure}
	\centering
	\label{fig:rev}
	\subfloat[]{%
		\resizebox{0.41\columnwidth}{!}{\definecolor{mycolor0}{rgb}{0,0,0}%
\definecolor{mycolor1}{rgb}{0.00000,0.44700,0.74100}%
\definecolor{mycolor2}{rgb}{0.85000,0.32500,0.09800}%
\definecolor{mycolor3}{rgb}{0.49400,0.18400,0.55600}%
\definecolor{mycolor4}{rgb}{0.92900,0.69400,0.12500}%
\definecolor{mycolor5}{rgb}{0.20000,0.60000,0.50000}%
\begin{tikzpicture}
\pgfplotsset{
    scale only axis,
    xmin=1, xmax=24,
    xtick={  2,  4,  6,  8,  10, 12, 14, 16, 18, 20, 22, 24},
      label style={font=\large},
ticklabel style={font=\large},
}
\pgfplotsset{every axis y label/.append style={mycolor0},
             every y tick label/.append style={mycolor0},
             y axis  line style={mycolor0}}
\begin{axis}[
  axis y line*=left,
  grid=both,
  ymin=50, ymax=450,
  xlabel=Time~(hr),
  ylabel= Total Revenue~(\pounds),
label style={font=\large},
ticklabel style={font=\large},
]
\addplot [color=mycolor2, line width=6.5pt]
  coordinates{
(1,376.996)
(2,402.731)
(3,418.0656)
(4,420.9572)
(5,390.4614)
(6,336.7156)
(7,286.3848)
(8,244.09)
(9,215.0792)
(10,211.3264)
(11,220.9794)
(12,216.2872)
(13,208.1232)
(14,192.4968)
(15,170.5498)
(16,140.03584)
(17,130.13304)
(18,137.27884)
(19,153.46484)
(20,182.73704)
(21,232.5424)
(22,281.257)
(23,313.910666666667)
(24,369.159)
};\label{plot_one}

\addplot [color=mycolor1, line width=2.0pt, mark=*, mark size=2.5, mark options={solid, mycolor1, fill=white}]
  coordinates{
(1,376.98675732)
(2,402.724840013333)
(3,418.061447966)
(4,420.953859866)
(5,390.460449216)
(6,336.680597312)
(7,286.349999496)
(8,244.055503144)
(9,215.045662606)
(10,211.291308446)
(11,220.977985754)
(12,216.28505658)
(13,208.120913324)
(14,192.402068632)
(15,170.428101944)
(16,139.91439461)
(17,130.012353947)
(18,137.114501877)
(19,153.388062851)
(20,182.688727335)
(21,232.494004901)
(22,281.206851414)
(23,313.908176843333)
(24,369.1501247)
};\label{plot_two}

\addplot [color=mycolor4, line width=2.0pt, mark=square, mark size=2.5, mark options={solid, mycolor4}]
  coordinates{
(1,315.1)
(2,326.911666666667)
(3,329.6336)
(4,329.7778)
(5,292.3224)
(6,243.7874)
(7,198.9756)
(8,161.1734)
(9,131.3162)
(10,145.08)
(11,155.5632)
(12,152.0348)
(13,157.579)
(14,148.7878)
(15,122.8362)
(16,93.0472)
(17,85.7778)
(18,78.0464)
(19,81.6014)
(20,96.152)
(21,139.8564)
(22,194.0192)
(23,238.984)
(24,347.462)
}; \label{plot_three}

\addplot [color=mycolor3, line width=2.0pt, mark=diamond, mark size=2.5, mark options={solid, mycolor3}]
  coordinates{
(1,335.703)
(2,346.429333333333)
(3,345.4218)
(4,349.4124)
(5,315.58)
(6,267.09074)
(7,229.07372)
(8,190.69632)
(9,153.21052)
(10,150.90872)
(11,166.09438)
(12,164.54946)
(13,159.06594)
(14,144.98082)
(15,121.11264)
(16,90.18308)
(17,79.01804)
(18,83.71336)
(19,98.61628)
(20,124.94346)
(21,171.28462)
(22,219.593)
(23,255.287666666667)
(24,320.114)
}; \label{plot_four}

\end{axis}

\pgfplotsset{every axis y label/.append style={mycolor5},
             every y tick label/.append style={mycolor5},
             y axis  line style={mycolor5}}
\begin{axis}[
  axis y line*=right,
  axis x line=none,
  ymin=1, ymax=5,
  ytick={  0,  0.5,  1,  1.5,  2.0,  2.5, 3.0, 3.5, 4.0, 4.5, 5.0},
  ylabel= Normalized Traffic Load,
  legend style={at={(0.385,1.0)}, anchor=north, legend cell align=left, align=left, draw=white!15!black},
  label style={font=\large},
ticklabel style={font=\large},
legend style={font=\large}
]
\addlegendimage{/pgfplots/refstyle=plot_one}\addlegendentry{ES}
\addlegendimage{/pgfplots/refstyle=plot_two}\addlegendentry{SA}
\addlegendimage{/pgfplots/refstyle=plot_three}\addlegendentry{A-type}
\addlegendimage{/pgfplots/refstyle=plot_four}\addlegendentry{D-type}
\addplot [color=mycolor5, line width=2.0pt, mark=asterisk, mark size=2.5, mark options={solid, mycolor5}]
  coordinates{
(1,1.85576)
(2,1.57309333333333)
(3,1.404692)
(4,1.372954)
(5,1.707134)
(6,2.295118)
(7,2.845924)
(8,3.307798)
(9,3.62423)
(10,3.663248)
(11,3.558762)
(12,3.609162)
(13,3.700176)
(14,3.87052)
(15,4.108582)
(16,4.437134)
(17,4.541538)
(18,4.462836)
(19,4.287666)
(20,3.971384)
(21,3.431688)
(22,2.904164)
(23,2.54731666666667)
(24,1.94177)
}; \addlegendentry{$\tau_{i,1}$}
\end{axis}

\end{tikzpicture}}\label{fig:rev_fxd}}
	\subfloat[]{%
		\resizebox{0.4\columnwidth}{!}{
%
%
\definecolor{mycolor1}{rgb}{0.00000,0.44700,0.74100}%
\definecolor{mycolor2}{rgb}{0.85000,0.32500,0.09800}%
\definecolor{mycolor3}{rgb}{0.92900,0.69400,0.12500}%
\definecolor{mycolor4}{rgb}{0.49400,0.18400,0.55600}%
\begin{tikzpicture}

\begin{axis}[%
width=4.521in,
height=3.566in,
at={(0.758in,0.481in)},
scale only axis,
xmin=1,
xmax=24,
xlabel style={font=\large},
xlabel={Time (hr)},
ymin=50,
ymax=400,
ylabel style={font=\large},
ylabel={Total Revenue (\pounds)},
axis background/.style={fill=white},
xmajorgrids,
xminorgrids,
ymajorgrids,
yminorgrids,
legend style={at={(0.85,1.0)}, anchor=north, legend cell align=left, align=left, draw=white!15!black},
label style={font=\large},
ticklabel style={font=\large},
legend style={font=\large}
]
\addplot [color=mycolor2, line width=6.5pt]
  table[row sep=crcr]{%
1	341.717\\
2	352.766666666667\\
3	360.3324\\
4	361.5\\
5	339.6322\\
6	300.1822\\
7	266.7868\\
8	240.2974\\
9	223.1246\\
10	223.9822\\
11	235.9964\\
12	231.7128\\
13	223.0018\\
14	206.8448\\
15	183.362\\
16	149.86034\\
17	138.78214\\
18	145.40994\\
19	160.77894\\
20	189.70594\\
21	238.029\\
22	281.675\\
23	308.636\\
24	350.227\\
};
\addlegendentry{ES}

\addplot [color=mycolor1, line width=2.0pt, mark=*, mark size=2.5, mark options={solid, mycolor1, fill=white}]
  table[row sep=crcr]{%
1	341.6955436\\
2	352.746716903333\\
3	360.312845252\\
4	361.478460362\\
5	339.612320376\\
6	300.160763454\\
7	266.763771866\\
8	240.25519761\\
9	223.084184938\\
10	223.939952062\\
11	235.953866914\\
12	231.65170362\\
13	222.959861188\\
14	206.79842205\\
15	183.246017086\\
16	149.294424122\\
17	138.234065962\\
18	144.846770764\\
19	160.2202937\\
20	189.214506934\\
21	237.986344472\\
22	281.632674794\\
23	308.605841753333\\
24	350.20272859\\
};
\addlegendentry{SA}

\addplot [color=mycolor3, line width=2.0pt, mark=square, mark size=2.5, mark options={solid, mycolor3}]
  table[row sep=crcr]{%
1	285.728\\
2	286.418\\
3	284.3104\\
4	283.3194\\
5	252.6878\\
6	215.7292\\
7	183.9594\\
8	157.0764\\
9	136.3084\\
10	153.7322\\
11	166.1888\\
12	162.9248\\
13	168.8854\\
14	159.8498\\
15	132.0544\\
16	99.6424\\
17	91.4998\\
18	82.7402\\
19	85.4254\\
20	99.8034\\
21	142.583\\
22	192.4636\\
23	233.131\\
24	329.665\\
};
\addlegendentry{A-type}

\addplot [color=mycolor4, line width=2.0pt, mark=diamond,  mark size=2.5, mark options={solid, mycolor4}]
  table[row sep=crcr]{%
1	304.222\\
2	303.811333333333\\
3	298.2844\\
4	300.5726\\
5	274.7542\\
6	237.32208\\
7	211.73636\\
8	186.11336\\
9	158.72096\\
10	160.07576\\
11	177.44668\\
12	176.32648\\
13	170.49438\\
14	155.80766\\
15	130.21006\\
16	96.52168\\
17	84.3059\\
18	88.5574\\
19	103.17492\\
20	129.63272\\
21	175.0191\\
22	219.2886\\
23	250.477333333333\\
24	304.053\\
};
\addlegendentry{D-type}

\end{axis}
\end{tikzpicture}%
}\label{fig:rev_NDT}}
		
	\subfloat[]{%
		\resizebox{0.41\columnwidth}{!}{\definecolor{mycolor0}{rgb}{0,0,0}%
\definecolor{mycolor1}{rgb}{0.00000,0.44700,0.74100}%
\definecolor{mycolor2}{rgb}{0.85000,0.32500,0.09800}%
\definecolor{mycolor3}{rgb}{0.49400,0.18400,0.55600}%
\definecolor{mycolor4}{rgb}{0.92900,0.69400,0.12500}%
\definecolor{mycolor5}{rgb}{0.20000,0.60000,0.50000}%
\begin{tikzpicture}
\pgfplotsset{
    scale only axis,
    xmin=1, xmax=24,
    xtick={  2,  4,  6,  8,  10, 12, 14, 16, 18, 20, 22, 24},
      label style={font=\large},
ticklabel style={font=\large},
}
\pgfplotsset{every axis y label/.append style={mycolor0},
             every y tick label/.append style={mycolor0},
             y axis  line style={mycolor0}}
\begin{axis}[
  axis y line*=left,
  grid=both,
  ymin=50, ymax=550,
  xlabel=Time~(hr),
  ylabel= Total Revenue~(\pounds),
label style={font=\large},
ticklabel style={font=\large},
]
\addplot [color=mycolor2, line width=6.5pt]
  coordinates{
(1,478.548)
(2,523.177333333333)
(3,543.0314)
(4,544.7772)
(5,501.2098)
(6,425.228)
(7,354.4038)
(8,293.9018)
(9,251.9848)
(10	244.475)
(11,251.7552)
(12,239.0704)
(13,223.017)
(14,197.9362)
(15,164.76006)
(16,127.33072)
(17,113.32286)
(18,115.61646)
(19,132.56026)
(20,171.1446)
(21,232.99754)
(22,301.1864)
(23,361.647666666667)
(24,443.694)
};\label{plot_one}

\addplot [color=mycolor1, line width=2.0pt, mark=*, mark size=2.5, mark options={solid, mycolor1, fill=white}]
  coordinates{
(1,478.52738174)
(2,523.1581178)
(3,543.012906776)
(4,544.75895707)
(5,501.19273679)
(6,425.209458396)
(7,354.38251505)
(8,293.881336188)
(9,251.963848756)
(10,244.01217043)
(11,251.292394436)
(12,238.607877132)
(13,222.554399118)
(14,196.895358606)
(15,163.9206677708)
(16,125.7160101464)
(17,111.7071186596)
(18,113.3611515016)
(19,130.4480803916)
(20,168.8802188328)
(21,231.5074185692)
(22,299.698312516)
(23,360.973177126667)
(24,443.6773761)
};\label{plot_two}

\addplot [color=mycolor4, line width=2.0pt, mark=square, mark size=2.5, mark options={solid, mycolor4}]
  coordinates{
(1,323.941)
(2,353.304333333333)
(3,369.2068)
(4,369.365)
(5,322.9322)
(6,258.485)
(7,183.5808)
(8,145.2634)
(9,112.4492)
(10,110.1028)
(11,122.42788)
(12,122.14372)
(13,105.03112)
(14,90.71102)
(15,80.80004)
(16,59.87196)
(17,60.92712)
(18,68.88292)
(19,79.67802)
(20,106.0976)
(21,149.7474)
(22,195.524)
(23,240.463333333333)
(24,293.773)
}; \label{plot_three}

\addplot [color=mycolor3, line width=2.0pt, mark=diamond, mark size=2.5, mark options={solid, mycolor3}]
  coordinates{
(1,432.148)
(2,431.466333333333)
(3,450.5318)
(4,459.4782)
(5,432.3954)
(6,366.7772)
(7,312.1826)
(8,	257.202)
(9,208.9874)
(10,196.745)
(11,209.6266)
(12,198.29276)
(13,175.68256)
(14,142.70712)
(15,109.4321)
(16,70.61588)
(17,57.3766)
(18,60.9672)
(19,72.37004)
(20,102.78626)
(21,149.57708)
(22,205.9196)
(23,257.73)
(24,345.669)
}; \label{plot_four}

\end{axis}

\pgfplotsset{every axis y label/.append style={mycolor5},
             every y tick label/.append style={mycolor5},
             y axis  line style={mycolor5}}
\begin{axis}[
  axis y line*=right,
  axis x line=none,
  ymin=2.1, ymax=3.1,
  ytick={2.1,  2.2,  2.3,  2.4,  2.5, 2.6, 2.7, 2.8, 2.9, 3.0, 3.1},
  ylabel= Normalized DT Spectrum Demand,
  legend style={at={(0.78,1.0)}, anchor=north, legend cell align=left, align=left, draw=white!15!black},
  label style={font=\large},
ticklabel style={font=\large},
legend style={font=\large}
]
\addlegendimage{/pgfplots/refstyle=plot_one}\addlegendentry{ES}
\addlegendimage{/pgfplots/refstyle=plot_two}\addlegendentry{SA}
\addlegendimage{/pgfplots/refstyle=plot_three}\addlegendentry{A-type}
\addlegendimage{/pgfplots/refstyle=plot_four}\addlegendentry{D-type}
\addplot [color=mycolor5, line width=2.0pt, dashed, mark=asterisk, mark size=2.5, mark options={solid, mycolor5}]
  coordinates{
(1,3.06888)
(2,3.03868666666667)
(3,3.018196)
(4,2.985566)
(5,2.965566)
(6,2.912768)
(7,2.847584)
(8,2.792994)
(9,2.729132)
(10,2.6589)
(11,2.610314)
(12,2.568418)
(13,2.492)
(14,2.416726)
(15,2.340522)
(16,2.248418)
(17,2.15562)
(18,2.119194)
(19,2.144128)
(20,2.265192)
(21,2.459584)
(22,2.660586)
(23,2.94643666666667)
(24,3.01447)
}; \addlegendentry{ $\Psi_{i,j}$~(DT)}
\end{axis}

\end{tikzpicture}}\label{fig:rev_DT}}
	\caption{The revenue obtained by the PN due to cell switching and spectrum leasing from fixed pricing policy and dynamic pricing policy~(DT and NDT spectrum demand) for 12 SBSs over a 24 hour period. (a) The left y-axis is the total revenue obtained from fixed pricing policy with NDT spectrum demand using ES, SA, A-type and D-type methods  while the right y-axis is the traffic load of the PN MBS. (b) Total revenue from dynamic pricing policy with NDT spectrum demand. (c) The left y-axis is the total revenue from dynamic pricing policy with DT spectrum demand using ES, SA, A-type and D type methods while the right y-axis is the traffic demand of the SN.}\label{fig:rev_total}
\end{figure}

Fig.~\ref{fig:rev_fxd} shows the hourly total revenue obtained by the PN following the fixed electricity and spectrum pricing policy with NDT spectrum demand using the proposed and benchmark methods. In addition, the traffic load of the PN MBS, $\tau_{i,1}$, is also presented.
The first thing we observe from Fig.~\ref{fig:rev_fxd} is that the revenue obtained from all methods follows a trend that is opposite of that of the traffic demanded of the PN. This is so because during the periods of the day where the PN traffic is low, more SBSs can be switched off which translates to more revenue generation from energy savings and spectrum leasing. The opposite is the case when the traffic of the PN is high.
Second, the SA algorithm follows ES almost exactly, since it is able to employ its mechanisms such as feasibility check and neighbourhood structures to determine the optimal cell switching and spectrum leasing pattern, but with much lesser complexity.

Third, both the A-Type and D-type heuristic solutions never outperform ES and SA algorithms because they also respect the constraint of not exceeding the MBS capacity. Even though they both respect the MBS capacity in order to maintain the QoS of the network, they utility, $\mathcal{N}$, used in determining which BSs to switch off only considers the difference in traffic demand between the PN and SN, but is not able to distinguish between the various types of BSs present. In this work, the PN and SN BSs have different capacities and power consumption, as a result, switching off a SBS with higher capacity and power consumption and leasing its spectrum to the SN would result in higher revenue than switching off one with a lower capacity. This limitation accounts for the lesser revenue obtained from both the A-type and D-type heuristics.

Another interesting point to discuss about the observations in Fig.~\ref{fig:rev_fxd} is that the D-type heuristic mostly outperforms the A-type heuristic because it switches off the SBS with highest utility, $\mathcal{N}$, values first and this helps in the generation of more revenue compared to A-type which does the opposite. However, this performance difference is mostly observable during the times of low traffic as there are more options and the higher utility is able to find a better solution. For the time when the network traffic is high, they start performing alike, since the number of cell switching and spectrum leasing options becomes very low. Overall, the performance difference between the D-type and A-type solutions is not large, as the former outperforms the later with a minimum of 1\% and a maximum of 29\%. 
The last observation worth discussing is that the SA solution mainly outperforms both A-type and D-type solutions~(by about 90\% and 65\% respectively) during periods of high traffic. The reason for this is that the number of cell switching and spectrum leasing options becomes very few during this period, thereby making it very difficult for them to find the best solution while the SA solution is carefully designed to be able to perform excellently well even in such periods.

Fig.~\ref{fig:rev_NDT} presents the total revenue obtained every hour by the PN when the dynamic pricing policy with NDT spectrum demand is considered using the proposed and benchmark methods. In the dynamic pricing policy, the prices of both electricity and spectrum vary at different times of the day depending on the amount of spectrum or electricity demanded.
Similar to what was observed in Fig.~\ref{fig:rev_fxd}, the pattern of the total revenue over the whole day is the inverse of the traffic profile of the PN. Moreover, the revenue is generally scaled down compared to Fig.~\ref{fig:rev_fxd}, and this is more noticeable during periods of low traffic. This is because a dynamic pricing policy is used, where the PN sometimes needs to lease the spectrum for less and at those times it also earns less from energy savings because the prices are lower. 
The D-type heuristic also slightly outperforms the A-type heuristics with almost the same percentage~(1\% to 29\%) as in Fig.~\ref{fig:rev_fxd}, due to the fact that higher utility values are considered first during cell switching which helps in greater revenue generation in the former compared to the later. The aforementioned confirms our previous argument on why the performance of the two benchmark algorithms are similar.
The proposed SA algorithm also greatly outperforms the A-type and D-type algorithms with a similar percentage~(90\% and 65\% respectively) as in Fig.~\ref{fig:rev_fxd} mostly during the period of high traffic in the PN because there are lesser cell switching and spectrum leasing options which make it difficult for the benchmark solutions to make the optimum decisions.
 
Fig.~\ref{fig:rev_DT} presents the total revenue obtained by the PN when dynamic pricing policy with DT spectrum demand is considered. In this case, the SN decides to delay its data transmission to periods when the spectrum price is low~(which also coincides with period of low traffic demand in the PN) so that they can access more spectrum at a cheaper rate. 
It can be observed that there is an overall increase in the total revenue obtained by the PN in Fig.~\ref{fig:rev_DT}, compared to Fig.~\ref{fig:rev_fxd} and Fig.~\ref{fig:rev_NDT}: the total revenue obtained from the proposed SA framework is about 19\% and 16\% higher than that obtained in the Fig.~\ref{fig:rev_fxd} and Fig.~\ref{fig:rev_NDT}, also it is evidenced by the peak value of the revenue of Fig.~\ref{fig:rev_DT} being about \pounds 183 and \pounds 123 higher than that in Fig.~\ref{fig:rev_fxd} and Fig.~\ref{fig:rev_NDT} respectively. This is because in dynamic pricing policy with DT spectrum demand, the SN can lease more spectrum as the periods of low traffic in the PN matches the period of high spectrum demand by the SN although the prices are lower. This statement is validated by comparing the traffic demand of the PN in Fig.~\ref{fig:rev_fxd} with the DT spectrum demand in Fig.~\ref{fig:rev_DT}; i.e., periods of lowest traffic demand in the PN~(e.g., in the first quarter of the day where the traffic load is 14\%) coincides with periods of highest spectrum demand from the SN~(about 28\%) so that even though the spectrum prices are lower at these times as seen in Fig.~\ref{fig:dyn_pri}, the large amount of spectrum demanded by the SN causes the total revenue in this scenario to be highest.

The performance difference between the D-type and A-type heuristics is more significant in this scenario compared to the fixed and dynamic pricing policy with NDT scenarios in Fig.~\ref{fig:rev_fxd} and Fig.~\ref{fig:rev_NDT} with values ranging from 5.3\% to 86\%. The reason for the wider performance gap is that the NDT spectrum demand is responsible for preventing the D-type heuristic from significantly outperforming A-type heuristic. This phenomenon originates from the fact that in the fixed and dynamic pricing policy with NDT spectrum demand, the trend of the SN traffic demand follows the PN traffic demand, hence the margin in the values of $\mathcal{N}$ is smaller in the both cases compared to the dynamic pricing policy with DT spectrum demand, thus accounting for the lesser total revenue results of A-type and D-type heuristics in the previous scenarios. 
On the other hand, for the dynamic pricing policy with DT spectrum demand, since the traffic demand of the SN is the inverse of the traffic load of the PN, the difference in the values of $\mathcal{N}$ at different time slots is higher and since D-type gives preference to SBSs with higher $\mathcal{N}$ during cell switching and spectrum leasing, more revenue is generated by the D-type compared to A-type, hence the reason for the wider margin in the revenue generated in the former compared to the later.
In addition, the SA algorithm greatly outperforms the A-type and D-type benchmarks in terms of revenue generation by 124\% and 95\% respectively. These values are 34\% and 31\% higher than its performance against the two benchmarks in both the fixed and dynamic pricing with NDT spectrum demand in Fig.~\ref{fig:rev_fxd} and Fig.~\ref{fig:rev_NDT} respectively. The reason is that the SA is able to take advantage of the more available options to switch off SBSs during period of low traffic which coincides with high spectrum demand by the SN in order to generate much higher revenue than in the two previous scenarios. 

It can also be observed that the results of D-type and A-type methods are almost the same in a few instances with the A-type even slightly surpassing that of the D-type at some points. 
For the NDT cases~(with both fixed and dynamic pricing policies) this occurs when both the data traffic of the PN and the spectrum demand of the SN are high. This is due to the fact that the difference in the values of the utility in this period is very small, thus, there is very little revenue from spectrum leasing as the SN is not able to access spectrum due to lack of dormant spectrum from the PN. Also, very little revenue can be obtained from energy saving since only very few SBSs can be turned off due to very high traffic load in the PN. For the DT case with dynamic pricing policy, the similarity in the results of both the D-type and A-type heuristics occur when the traffic demand of the PN is high and the spectrum demand of the SN is low. At these periods, both benchmarks begin to function alike because even though there is a large difference in the value of the utility function, there is very little opportunity to switch off the SBSs due to high traffic in the PN. Hence, there is an insignificant difference in the performance of both benchmarks as relatively less revenue can be obtained during this period.


\begin{figure}[h!]
	\centering
	\subfloat[Total amount spent on spectrum purchase and the quantity of spectrum obtained by SN for a 24 hour period.]{%
	\resizebox{0.45\columnwidth}{!}{
%
%
\definecolor{mycolor1}{rgb}{0.92900,0.69400,0.12500}%
\definecolor{mycolor2}{rgb}{0.46600,0.67400,0.18800}%
\definecolor{mycolor5}{rgb}{0.20000,0.60000,0.50000}%
\begin{tikzpicture}
\pgfplotsset{every axis y label/.append style={mycolor2, font=\Large},
             every y tick label/.append style={mycolor2, font=\Large},
             y axis  line style={mycolor2}}

\begin{axis}[%
width=4.521in,
height=3.566in,
at={(0.758in,0.481in)},
scale only axis,
bar shift auto,
log origin=infty,
xmin=0.514285714285714,
xmax=4.48571428571429,
xtick={1,2,3,4},
xticklabels={{ES~(DT)},{SA~(DT)},{ES~(NDT)},{SA~(NDT)}},
ymin=4500,
ymax=7500,
ylabel={Total expenditure~(\pounds)},
axis background/.style={fill=white},
xmajorgrids,
xminorgrids,
ymajorgrids,
yminorgrids,
legend style={at={(0.78,1.0)}, anchor=north, legend cell align=left, align=left, draw=white!15!black},
label style={font=\large},
ticklabel style={font=\large},
legend style={font=\large}
]
\addplot[ybar, bar width=0.4, fill=mycolor2, draw=black, area legend] table[row sep=crcr] {%
1	7199.7\\
2	7182.2\\
3	6028.9\\
4	6025.9\\
};
\addlegendentry{Expenditure}

\addplot[ybar, bar width=0.4, fill=mycolor1, draw=black, area legend] table[row sep=crcr] {%
1	5831.6\\
2	5818.8\\
3	4791.3\\
4	4789.2\\
};
\addlegendentry{Spectrum Purchased}

\end{axis}
\pgfplotsset{every axis y label/.append style={mycolor1, font=\Large},
             every y tick label/.append style={mycolor1, font=\Large},
             y axis  line style={mycolor1}}

    \begin{axis}[
        width=4.521in,
height=3.566in,
at={(0.758in,0.481in)},
scale only axis,
bar shift auto,
log origin=infty,
        ymin=4500,
        ymax=7500,
        axis x line=none,
        axis y line*=right,
        ylabel={Total spectrum purchased~(RBs)},
        bar shift={\pgfplotbarwidth/2},
label style={font=\large},
ticklabel style={font=\large},
    ]
\addplot[ybar, bar width=0.4, fill=mycolor1, draw=black, area legend] table[row sep=crcr] {%
1	5831.6\\
2	5818.8\\
3	4791.3\\
4	4789.2\\
};
    \end{axis}
\end{tikzpicture}%
}\label{fig:max_rev}}
	\subfloat[Unit cost per RB.] {%
	\resizebox{0.38\columnwidth}{!}{
%
%
\definecolor{mycolor1}{rgb}{0.00000,0.44700,0.74100}%
\definecolor{mycolor2}{rgb}{0.85000,0.32500,0.09800}%
\begin{tikzpicture}

\begin{axis}[%
width=4.521in,
height=3.566in,
at={(0.758in,0.481in)},
scale only axis,
bar shift auto,
log origin=infty,
xmin=-0.2,
xmax=5.2,
xtick={0.6,1.6,3.5,4.5},
xticklabels={{ES},{SA},{ES},{SA}},
ymin=1.23,
ymax=1.26,
ylabel style={font=\color{white!15!black}},
ylabel={Unit cost of Spectrum (\pounds)},
axis background/.style={fill=white},
xmajorgrids,
xminorgrids,
ymajorgrids,
yminorgrids,
legend style={at={(0.09,1.0)}, anchor=north, legend cell align=left, align=left, draw=white!15!black},
label style={font=\large},
ticklabel style={font=\large},
legend style={font=\large}
]
\addplot[ybar, bar width=0.8, fill=mycolor1, draw=black, area legend] table[row sep=crcr] {%
1	1.2346\\
2	1.2343\\
};
\addlegendentry{DT}

\addplot[ybar, bar width=0.8, fill=mycolor2, draw=black, area legend] table[row sep=crcr] {%
3	1.2583\\
4	1.2582\\
};
\addlegendentry{NDT}

\end{axis}
\end{tikzpicture}%
}\label{fig:uni_cst}}
	\caption{Total expenditure and quantity of spectrum purchased by SN and the average unit cost of the spectrum for 12 SBSs.}
\end{figure}
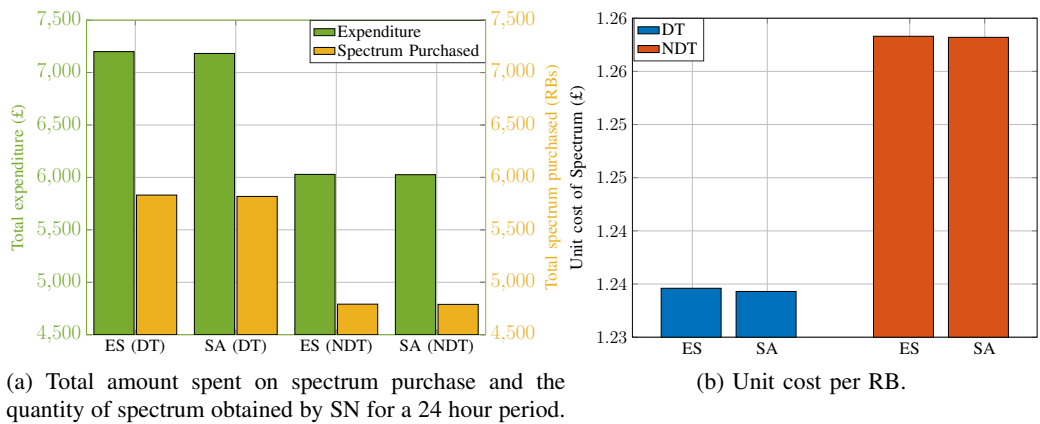

Fig.~\ref{fig:max_rev} shows the total amount spent by the SN for spectrum purchase as well as the total quantity of spectrum obtained for a 24 hours period using the proposed SA-based framework and ES while Fig.~\ref{fig:uni_cst} shows the unit cost of the spectrum~(i.e., price per RB) for both DT and NDT spectrum demand using both algorithms.
From Fig.~\ref{fig:max_rev} we can see that the total amount expended by the SN on spectrum purchase as well as the quantity of spectrum purchased are significantly higher in DT than in the NDT scenario with a percentage difference of 19\% and 21\% respectively. 
The rationale behind this is that most of the periods when the electricity and spectrum prices are low are also the periods when the traffic loads of the MBS and SBSs are low. As such, more SBSs can be turned off in order to ensure that more spectrum is available for SN to purchase during these periods. 
Although more spectrum is available to the SN for both DT and NDT spectrum leasing scenarios with the dynamic pricing polices during periods of low traffic load in the PN, the difference in the volume of spectrum demanded in both cases is what accounts for the difference in the amount expended on spectrum purchase in Fig.~~\ref{fig:max_rev}. 

In the DT case, the data to be transmitted is delayed until when the spectrum and electricity prices are low, which means that the SN is able to take advantage of more spectrum available in order to offer more data services to it users. However, for the NDT case, even though more spectrum is available during periods of low prices, the spectrum demanded by the SN during this period is also low, so lesser revenue is generated and fewer data services can be offered in this scenario.
For example, in the first quarter of the day where the traffic load of the PN is the lowest~(about 14\%), the revenue generated by the dynamic pricing policy with DT spectrum demand is 47\% higher than that obtained from the dynamic spectrum demand with NDT spectrum demand because more spectrum is available for leasing as well as a corresponding high spectrum demand from the SN. However, the availability of more spectrum does not correspond with high spectrum demand in the NDT case thereby leading to a lesser revenue generation.
Fig.~\ref{fig:max_rev} also reveals that the total expenditure and quantity of spectrum purchased using SA algorithm is almost the same as that of ES algorithm which validates the excellent performance of the SA algorithm earlier discussed under Fig.~\ref{fig:rev_fxd}, Fig.~\ref{fig:rev_NDT} and Fig.~\ref{fig:rev_DT}.

The purchase of more spectrum by the SN in the DT case compared to the NDT case means that the SN incurs more expenses during DT data transmission compared to NDT data transmission. Therefore, the DT case is more beneficial to the PN because it results in more total revenue. It is also beneficial to the SN because it pays less for a unit of spectrum even though its total expenditures increases.
Hence, where possible (for suitable applications), we can conclude that the shift in the SN traffic demand would be recommended. However, the kind of shift in the data transmission time of the SN does not have to be implemented in exactly the same way as in this work, instead, depending on the type of application, the latency requirements are evaluated and the appropriate shifts in the traffic is implemented accordingly, making the DT spectrum demand quite flexible and dynamic. Although this traffic shift may not always coincide with the cheapest time but to a cheaper time.
In summary, DT spectrum demand will make the business of both PNs and SNs more sustainable because it is more profitable for both parties.

A major constraint in this work is to ensure that the QoS of the network is maintained by ensuring that traffic served by the network remains constant even when some SBSs are switched off. The PN is supposed to respect the capacity constraints of the MBS before switching off any SBS. From the simulations, we observed that both the proposed and benchmark solutions are able to maintain the QoS of the PN. The SA algorithm uses the feasibility check in Algorithm~\ref{alg:Fes_chk} to ensure that only solutions that do not exceed the capacity of the MBS are considered. The ES algorithm follows similar procedure by guaranteeing that solutions that exceeds the MBS capacity are excluded when selecting the optimal cell switching and spectrum leasing strategy. Both the A-type and D-type algorithms are implemented in such a way that the traffic load of the SBSs are offloaded sequentially~(in ascending and descending order respectively) and once the offloading capacity of the MBS is attained, no further SBS is turned off. By so doing, they both guarantee that the throughput of the network is maintained.
It is also worthy of note that irrespective of the pricing model used for electricity and spectrum~(fixed or dynamic) and the type of spectrum demanded by the SN~(DT or NDT), the average throughput of the PN remains the same. This is because both the proposed and benchmark algorithms take the traffic-QoS constraint~\eqref{eq:max_QoS} into consideration thereby ensuring that the QoS of the network is not violated.

\section{Conclusion}
In this paper, we considered the problem of revenue maximization through cell switching and spectrum leasing in order to maximize the revenue of the PN, which comprises a HetNet with different types of SBSs while the SN comprises SN BSs. An SA algorithm based solution was proposed to determine the optimal cell switching and spectrum leasing strategy that would result in maximum revenue for the PN while ensuring that the QoS of the network is maintained. We considered fixed and dynamic pricing policy for both electricity and spectrum. Under the dynamic pricing policy, both DT and NDT spectrum demand scenarios were considered in order to determine the effect of these policies on the revenue of the PN as well as the expenditure and amount of service demands that can be met by the SN. 

The simulation results show that the PN is able to obtain more revenue using the dynamic pricing policy with DT spectrum demand.
Moreover, in the DT spectrum demand scenario, the SN is able to lease more spectrum when the spectrum prices are low, which enables it to serve more data services at a reduced average unit price. Thus, making this scenario more profitable to the SN compared to the fixed or dynamic pricing policy with NDT spectrum demand scenarios. Overall, the performance of the proposed method is almost the same as that of the ES algorithm with lesser time complexity.
In future, we intend to investigate a scenario whereby more than one SN BSs would be competing to lease the spectrum of each PN BSs and the PN would have to decide which of the SN BSs to lease the spectrum as well as the possibility of leasing the spectrum to more than one SN-BSs at the same time.
\bibliographystyle{IEEEtran}
\bibliography{ref}

\begin{thebibliography}{10}
\providecommand{\url}[1]{#1}
\csname url@samestyle\endcsname
\providecommand{\newblock}{\relax}
\providecommand{\bibinfo}[2]{#2}
\providecommand{\BIBentrySTDinterwordspacing}{\spaceskip=0pt\relax}
\providecommand{\BIBentryALTinterwordstretchfactor}{4}
\providecommand{\BIBentryALTinterwordspacing}{\spaceskip=\fontdimen2\font plus
\BIBentryALTinterwordstretchfactor\fontdimen3\font minus
  \fontdimen4\font\relax}
\providecommand{\BIBforeignlanguage}[2]{{%
\expandafter\ifx\csname l@#1\endcsname\relax
\typeout{** WARNING: IEEEtran.bst: No hyphenation pattern has been}%
\typeout{** loaded for the language `#1'. Using the pattern for}%
\typeout{** the default language instead.}%
\else
\language=\csname l@#1\endcsname
\fi
#2}}
\providecommand{\BIBdecl}{\relax}
\BIBdecl

\bibitem{MORGADO2018}
A.~Morgado, K.~M.~S. Huq, S.~Mumtaz, and J.~Rodriguez, ``{A survey of 5G
  technologies: regulatory, standardization and industrial perspectives},''
  \emph{Digital Communications and Networks}, vol.~4, no.~2, pp. 87 -- 97,
  2018.

\bibitem{Akpakwu2018}
G.~A. {Akpakwu}, B.~J. {Silva}, G.~P. {Hancke}, and A.~M. {Abu-Mahfouz}, ``{A
  Survey on 5G Networks for the Internet of Things: Communication Technologies
  and Challenges},'' \emph{IEEE Access}, vol.~6, pp. 3619--3647, 2018.

\bibitem{Kamel2016}
M.~Kamel, W.~Hamouda, and A.~Youssef, ``Ultra-dense networks: A survey,''
  \emph{IEEE Communications Surveys Tutorials}, vol.~18, no.~4, pp. 2522--2545,
  2016.

\bibitem{Buzzi2016}
S.~{Buzzi}, C.~{I}, T.~E. {Klein}, H.~V. {Poor}, C.~{Yang}, and A.~{Zappone},
  ``{A Survey of Energy-Efficient Techniques for 5G Networks and Challenges
  Ahead},'' \emph{IEEE Journal on Selected Areas in Communications}, vol.~34,
  no.~4, pp. 697--709, 2016.

\bibitem{Giordani2020}
M.~Giordani, M.~Polese, M.~Mezzavilla, S.~Rangan, and M.~Zorzi, ``Toward {6G}
  networks: Use cases and technologies,'' \emph{IEEE Communications Magazine},
  vol.~58, no.~3, pp. 55--61, 2020.

\bibitem{Feng2017}
M.~{Feng}, S.~{Mao}, and T.~{Jiang}, ``{Base Station ON-OFF Switching in 5G
  Wireless Networks: Approaches and Challenges},'' \emph{IEEE Wireless
  Communications}, vol.~24, no.~4, pp. 46--54, 2017.

\bibitem{ALAMU2020}
\BIBentryALTinterwordspacing
O.~Alamu, A.~Gbenga-Ilori, M.~Adelabu, A.~Imoize, and O.~Ladipo, ``Energy
  efficiency techniques in ultra-dense wireless heterogeneous networks: An
  overview and outlook,'' \emph{Engineering Science and Technology, an
  International Journal}, vol.~23, no.~6, pp. 1308 -- 1326, 2020. [Online].
  Available:
  \url{http://www.sciencedirect.com/science/article/pii/S2215098619328745}
\BIBentrySTDinterwordspacing

\bibitem{Yua2018}
K.~A. {Yau}, J.~{Qadir}, C.~{Wu}, M.~A. {Imran}, and M.~H. {Ling},
  ``{Cognition-Inspired 5G Cellular Networks: A Review and the Road Ahead},''
  \emph{IEEE Access}, vol.~6, pp. 35\,072--35\,090, 2018.

\bibitem{Asad2019}
S.~M. {Asad}, M.~{Ozturk}, R.~N. {Bin Rais}, A.~{Zoha}, S.~{Hussain}, Q.~H.
  {Abbasi}, and M.~A. {Imran}, ``Reinforcement learning driven energy efficient
  mobile communication and applications,'' in \emph{2019 IEEE International
  Symposium on Signal Processing and Information Technology (ISSPIT)}, 2019,
  pp. 1--7.

\bibitem{Abubakar2020}
A.~I. {Abubakar}, M.~{Ozturk}, R.~N.~B. {Rais}, S.~{Hussain}, and M.~A.
  {Imran}, ``Load-aware cell switching in ultra-dense networks: An artificial
  neural network approach,'' in \emph{2020 International Conference on UK-China
  Emerging Technologies (UCET)}, 2020, pp. 1--4.

\bibitem{Ozturk2021}
M.~Ozturk, A.~I. Abubakar, J.~P.~B. Nadas, R.~N.~B. Rais, S.~Hussain, and M.~A.
  Imran, ``Energy optimization in ultra-dense radio access networks via
  traffic-aware cell switching,'' \emph{IEEE Transactions on Green
  Communications and Networking}, vol.~5, no.~2, pp. 832--845, 2021.

\bibitem{Zhang2020}
K.~{Zhang}, X.~{Wen}, Y.~{Chen}, and Z.~{Lu}, ``Deep reinforcement learning for
  energy saving in radio access network,'' in \emph{2020 IEEE/CIC International
  Conference on Communications in China (ICCC Workshops)}, 2020, pp. 35--40.

\bibitem{Wu2020}
S.~{Wu}, Y.~{Wang}, and L.~{Bai}, ``Deep convolutional neural network assisted
  reinforcement learning based mobile network power saving,'' \emph{IEEE
  Access}, vol.~8, pp. 93\,671--93\,681, 2020.

\bibitem{Amine2020}
A.~E. {Amine}, P.~{Dini}, and L.~{Nuaymi}, ``Reinforcement learning for
  delay-constrained energy-aware small cells with multi-sleeping control,'' in
  \emph{2020 IEEE International Conference on Communications (ICC) Workshops},
  2020, pp. 1--6.

\bibitem{Alsafasfeh2020}
Q.~Alsafasfeh, O.~A. Saraereh, A.~Ali, L.~Al-Tarawneh, I.~Khan, and A.~Silva,
  ``Efficient power control framework for small-cell heterogeneous networks,''
  \emph{Sensors}, vol.~20, no.~5, 2020.

\bibitem{Tan2018}
X.~J. {Tan} and W.~{Zhan}, ``Traffic-adaptive spectrum leasing between primary
  and secondary networks,'' \emph{IEEE Transactions on Vehicular Technology},
  vol.~67, no.~7, pp. 6546--6560, 2018.

\bibitem{Tsirakis2018}
C.~{Tsirakis}, E.~{Lopez-Aguilera}, P.~{Matzoros}, G.~{Agapiou}, and
  D.~{Varoutas}, ``{Spectrum Trading in Virtualized Multi-Tenant 5G
  Networks},'' in \emph{2018 15th International Symposium on Wireless
  Communication Systems (ISWCS)}, 2018, pp. 1--6.

\bibitem{Liu2018}
X.~{Liu}, L.~{Li}, W.~{Liang}, F.~{Yang}, H.~{Xu}, and Z.~{Han}, ``Joint
  optimization scheme for spectrum leasing in cognitive radio networks,'' in
  \emph{2018 10th International Conference on Wireless Communications and
  Signal Processing (WCSP)}, 2018, pp. 1--6.

\bibitem{Bilibashi2020}
\BIBentryALTinterwordspacing
D.~Bilibashi, E.~M. Vitucci, V.~Degli-Esposti, and A.~Giorgetti, ``An
  energy-efficient unselfish spectrum leasing scheme for cognitive radio
  networks,'' \emph{Sensors}, vol.~20, no.~21, 2020. [Online]. Available:
  \url{https://www.mdpi.com/1424-8220/20/21/6161}
\BIBentrySTDinterwordspacing

\bibitem{Liu2020}
Z.~{Liu}, M.~{Zhao}, K.~Y. {Chan}, Y.~{Yuan}, and X.~{Guan}, ``Approach of
  robust resource allocation in cognitive radio network with spectrum
  leasing,'' \emph{IEEE Transactions on Green Communications and Networking},
  vol.~4, no.~2, pp. 413--422, 2020.

\bibitem{Xiao2020}
\BIBentryALTinterwordspacing
X.~Xiao, F.~Zeng, Z.~Hu, and L.~Jiao, ``Dynamic flow-adaptive spectrum leasing
  with channel aggregation in cognitive radio networks,'' \emph{Sensors},
  vol.~20, no.~13, 2020. [Online]. Available:
  \url{https://www.mdpi.com/1424-8220/20/13/3800}
\BIBentrySTDinterwordspacing

\bibitem{Ozturk2019}
M.~{Ozturk}, A.~I. {Abubakar}, N.~U. {Hassan}, S.~{Hussain}, M.~A. {Imran}, and
  C.~{Yuen}, ``Spectrum cost optimization for cognitive radio transmission over
  tv white spaces using artificial neural networks,'' in \emph{2019 UK/ China
  Emerging Technologies (UCET)}, 2019, pp. 1--4.

\bibitem{Sboui2015}
N.~{Sboui}, H.~{Ghazzai}, Z.~{Rezki}, and M.~{Alouini}, ``Green collaboration
  in cognitive radio cellular networks with roaming and spectrum trading,'' in
  \emph{2015 IEEE 26th Annual International Symposium on Personal, Indoor, and
  Mobile Radio Communications (PIMRC)}, \NOOP{2015}2015, pp. 1420--1425.

\bibitem{Sboui2016}
L.~{Sboui}, H.~{Ghazzai}, Z.~{Rezki}, and M.~{Alouini}, ``On green cognitive
  radio cellular networks: Dynamic spectrum and operation management,''
  \emph{IEEE Access}, vol.~4, pp. 4046--4057, \NOOP{2016b}2016.

\bibitem{vassaki2015}
S.~{Vassaki}, M.~I. {Poulakis}, and A.~D. {Panagopoulos}, ``Spectrum leasing in
  cognitive radio networks: A matching theory approach,'' in \emph{2015 IEEE
  81st Vehicular Technology Conference (VTC Spring)}, 2015, pp. 1--5.

\bibitem{Mohamed2016}
A.~Mohamed, O.~Onireti, M.~A. Imran, A.~Imran, and R.~Tafazolli, ``Control-data
  separation architecture for cellular radio access networks: A survey and
  outlook,'' \emph{IEEE Communications Surveys Tutorials}, vol.~18, no.~1, pp.
  446--465, 2016.

\bibitem{Auer2011}
G.~{Auer}, V.~{Giannini}, C.~{Desset}, I.~{Godor}, P.~{Skillermark},
  M.~{Olsson}, M.~A. {Imran}, D.~{Sabella}, M.~J. {Gonzalez}, O.~{Blume}, and
  A.~{Fehske}, ``How much energy is needed to run a wireless network?''
  \emph{IEEE Wireless Communications}, vol.~18, no.~5, pp. 40--49, 2011.

\bibitem{Debaillie2015}
B.~{Debaillie}, C.~{Desset}, and F.~{Louagie}, ``A flexible and future-proof
  power model for cellular base stations,'' in \emph{2015 IEEE 81st Vehicular
  Technology Conference (VTC Spring)}, 2015, pp. 1--7.

\bibitem{EID2016elect_Pri}
C.~Eid, E.~Koliou, M.~Valles, J.~Reneses, and R.~Hakvoort, ``Time-based pricing
  and electricity demand response: Existing barriers and next steps,''
  \emph{Utilities Policy}, vol.~40, pp. 15--25, 2016.

\bibitem{3gpp2018technical}
3GPP, ``Technical specification group radio access network; {NR}; physical
  channels and modulation ({R}elease 15),'' 2018.

\bibitem{wei2018simulated}
L.~Wei, Z.~Zhang, D.~Zhang, and S.~C. Leung, ``A simulated annealing algorithm
  for the capacitated vehicle routing problem with two-dimensional loading
  constraints,'' \emph{European Journal of Operational Research}, vol. 265,
  no.~3, pp. 843--859, 2018.

\bibitem{Lin2018}
Z.~Lin, J.~Wang, Z.~Fang, M.~Hu, C.~Cai, and J.~Zhang, ``Accurate maximum power
  tracking of wireless power transfer system based on simulated annealing
  algorithm,'' \emph{IEEE Access}, vol.~6, pp. 60\,881--60\,890, 2018.

\bibitem{hansen2010variable}
P.~Hansen, N.~Mladenovi{\'c}, and J.~A.~M. P{\'e}rez, ``Variable neighbourhood
  search: methods and applications,'' \emph{Annals of Operations Research},
  vol. 175, no.~1, pp. 367--407, 2010.

\bibitem{YuNN}
V.~F. Yu, A.~P. Redi, Y.~A. Hidayat, and O.~J. Wibowo, ``A simulated annealing
  heuristic for the hybrid vehicle routing problem,'' \emph{Appl. Soft
  Comput.}, vol.~53, no.~C, p. 119–132, Apr. 2017.

\bibitem{alvarez2018iterated}
A.~Alvarez, P.~Munari, and R.~Morabito, ``Iterated local search and simulated
  annealing algorithms for the inventory routing problem,'' \emph{International
  Transactions in Operational Research}, vol.~25, no.~6, pp. 1785--1809, 2018.

\bibitem{chibante2010simulated}
R.~Chibante, \emph{Simulated annealing: theory with applications}.\hskip 1em
  plus 0.5em minus 0.4em\relax BoD--Books on Demand, 2010.

\bibitem{barlacchi2015multi}
G.~Barlacchi, M.~De~Nadai, R.~Larcher, A.~Casella, C.~Chitic, G.~Torrisi,
  F.~Antonelli, A.~Vespignani, A.~Pentland, and B.~Lepri, ``A multi-source
  dataset of urban life in the city of milan and the province of trentino,''
  \emph{Scientific data}, vol.~2, no.~1, pp. 1--15, 2015.

\end{thebibliography}
\end{document}